\documentclass[twocolumn]{aastex631}

\usepackage{hyperref}
\usepackage{booktabs}
\usepackage{multirow}
\usepackage{amsmath}

\begin{document}

\title{FRESCO: Constraining Dust Attenuation and Star-Formation Rates of $z\sim 2$ Star-Forming Galaxies with \textit{JWST} Paschen and Ground-Based Balmer Emission Line Observations}

\author[0000-0002-1033-3656]{Michael A. Wozniak}
\affiliation{Department of Physics and Astronomy, University of California, Riverside, 900 University Avenue, Riverside, CA 92521, USA}

\author[0000-0001-9687-4973]{Naveen A. Reddy}
\affiliation{Department of Physics and Astronomy, University of California, Riverside, 900 University Avenue, Riverside, CA 92521, USA}

\author[0000-0001-5851-6649]{Pascal A. Oesch}
\affiliation{Department of Astronomy, University of Geneva, Chemin Pegasi 51, 1290 Versoix, Switzerland}
\affiliation{Cosmic Dawn Center (DAWN), Niels Bohr Institute, University of Copenhagen, Jagtvej 128, K\o benhavn N, DK-2200, Denmark}

\author[0000-0001-5346-6048]{Ivan Kramarenko}
\affiliation{Institute of Science and Technology Austria (ISTA), Am Campus 1, 3400 Klosterneuburg, Austria}

\author[0000-0003-2871-127X]{Jorryt Matthee}
\affiliation{Institute of Science and Technology Austria (ISTA), Am Campus 1, 3400 Klosterneuburg, Austria}

\author[0000-0002-6558-9894]{Chloe Neufeld}
\affiliation{Department of Astronomy, Yale University, New Haven, CT 06520, USA}

\author[0000-0003-4702-7561]{Irene Shivaei}
\affiliation{Centro de Astrobiolog\'{i}a (CAB), CSIC-INTA, Carretera de Ajalvir km 4, Torrej\'{o}n de Ardoz, E-28850, Madrid, Spain}

\author[0000-0001-5846-4404]{Bahram Mobasher}
\affiliation{Department of Physics and Astronomy, University of California, Riverside, 900 University Avenue, Riverside, CA 92521, USA}

\author[0000-0003-3509-4855]{Alice E. Shapley}
\affiliation{Department of Physics and Astronomy, University of California, Los Angeles, 430 Portola Plaza, Los Angeles, CA 90095, USA}

\author[0000-0002-4935-9511]{Brian Siana}
\affiliation{Department of Physics and Astronomy, University of California, Riverside, 900 University Avenue, Riverside, CA 92521, USA}

\correspondingauthor{Michael A. Wozniak}
\email{michael.wozniak@email.ucr.edu}

\begin{abstract}

We present new constraints on dust attenuation and star-formation rates (SFRs) for 77 galaxies at redshifts $z=1.43-2.65$, using Paschen emission line detections from the \textit{JWST} FRESCO survey and ground-based Balmer line measurements from the MOSDEF survey. Using nebular and continuum emission maps, we find that Paschen emission covers a smaller area than continuum emission observed in the F210M (2.1\,$\mu$m; rest-frame optical) and F444W (4.4\,$\mu$m; rest-frame near-IR) bands, and is preferentially located toward galaxy outskirts. These results suggest that current star formation is concentrated in regions farther from galaxy centers than older stellar populations traced by the continuum, indicative of inside-out star formation. With a careful accounting of slit-loss corrections for ground-based measurements, we calculate nebular reddening and dust-corrected SFRs using the Balmer decrement (H$\alpha$/H$\beta$) and Paschen-to-Balmer line ratios (Pa$\alpha$/H$\alpha$ and Pa$\beta$/H$\alpha$), assuming the Milky Way extinction curve. On average, Paschen-derived reddening and SFRs agree with Balmer-derived values; however, two galaxies exhibit significantly higher Paschen reddening and four show significantly higher Paschen SFRs. We find that non-unity dust covering fractions bias the Balmer decrement toward less reddened OB associations, while decrements involving the Paschen lines are less affected by this bias. These results highlight the enhanced sensitivity of the Paschen lines to the most heavily obscured OB associations in $z\sim2$ galaxies, particularly in galaxies with patchy dust geometries. Future studies using Paschen lines exclusively to measure nebular reddening will yield more robust constraints on the dustiest star-forming regions.

\end{abstract}



\section{Introduction} \label{sec:intro}

Cosmic noon ($z\!\sim\!1\!-\!3$) is of particular interest for extragalactic studies, as it marks the epoch of peak star formation \citep{Madau2014}. The elevated star-formation rates (SFRs) of galaxies during this period coincide with a rapid buildup of interstellar dust \citep{Schreiber2020, Perrotta2024}. Since dust absorbs and scatters light --- effects that become more severe at shorter wavelengths --- it must be accounted for when inferring several fundamental properties of galaxies. These include the age and mass of the stellar population derived from the stellar continuum, as well as the SFR and metallicity inferred from nebular emission. Hydrogen recombination lines are commonly used to constrain the effects of nebular dust reddening and probe instantaneous SFRs \citep{Kennicutt1994, Kennicutt2012}. Originating from ionized (H\,$\rm\text{\small II}$) regions around massive, short-lived stars, these lines are robust tracers of recent star formation and are less sensitive to a galaxy's star-formation history compared to UV-based indicators \citep{Kennicutt1998}. In local galaxies, optical spectroscopy has been widely used to study nebular reddening and SFRs with recombination lines \citep[e.g.,][]{Cardelli1989, Calzetti1994, Johnson2007, Wild2011, Kreckel2013, Rezaee2021, Cleri2022, Gimenez2022, Maheson2024}, while near-IR spectroscopy has enabled similar studies at $z\!\sim\!1\!-\!3$ \citep[e.g.,][]{Schreiber2009, Kashino2013, Price2014, Reddy2015, Reddy2020, Shivaei2020, Battisti2022, Reddy2023, Shapley2023, Maheson2024, Seille2024}. Among these studies, the Balmer decrement (H$\alpha$/H$\beta$) has been a cornerstone for constraining nebular reddening in high-redshift galaxies.

The wavelength-dependent reddening effects of dust are described by a galaxy's attenuation curve, which accounts for both extinction (dust grains absorbing and scattering light out of the line of sight) and the scattering of light into the line of sight. Extinction depends primarily on the composition and size distribution of the dust grains, while the spatial arrangement of dust relative to the stars determines how light is scattered into the line of sight (\citeauthor{Salim2020} \citeyear{Salim2020} and references therein). The \textit{stellar} attenuation curve describes these effects on the stellar continuum. In high-redshift star-forming galaxies, the Small Magellanic Cloud \citep[SMC;][]{Gordon2003} and Calzetti \citep{Calzetti2000} curves are commonly assumed to describe stellar continuum reddening \citep[e.g.,][]{Schreiber2009, Yoshikawa2010, Wuyts2011, Price2014, Bouwens2016, Fudamoto2017, Reddy2018, Reddy2018_b, Theios2019, Reddy2020, Reddy2023}. Accurate dust corrections to the continuum are particularly important when using UV-based SFR indicators \citep{Hirashita2003}, as shorter-wavelength emission is more affected by attenuation. The \textit{nebular} attenuation curve describes the absorption and scattering of light along the lines of sight toward H\,$\rm\text{\small II}$ regions. Accurate dust corrections to nebular emission are critical when constraining galaxy properties using lines that are widely separated in wavelength, including hydrogen recombination lines for SFRs \citep{Kennicutt1994}, $\rm[O\,\text{\small III}]\lambda5007/[O\,\text{\small II}]\lambda\lambda3726,3729$ for the ionization parameter \citep{Kewley2002}, and $\rm([O\,\text{\small II}]\lambda\lambda3726,3729+[O\,\text{\small III}]\lambda\lambda4959,5007)/H\beta$ for R23 metallicity \citep{Pagel1979, Tremonti2004}. The Milky Way (MW) extinction curve \citep{Cardelli1989} is commonly assumed to describe the reddening of nebular emission in high-redshift star-forming galaxies \citep[e.g.,][]{Steidel2014, Reddy2015, Shivaei2015, Theios2019, Shivaei2020, Fetherolf2021, Lorenz2023, Reddy2023, Reddy2025}.

While the Balmer decrement has been pivotal in constraining nebular reddening and SFRs in high-redshift galaxies, it may fail to trace emission from optically thick (i.e., heavily dust-obscured) regions. Because of their longer wavelengths, the Pa$\alpha$ and Pa$\beta$ lines (the two lowest-order Paschen series lines) are less affected by dust attenuation relative to the Balmer lines. As a result, decrements using Pa$\alpha$ and Pa$\beta$ can probe nebular reddening and star formation in regions that are optically thick in the Balmer lines \citep{Gimenez2022, Prescott2022, Reddy2023, Lorenz2025, Reddy2025}. In addition, nebular attenuation curves exhibit less variation in shape at the wavelengths of the Paschen lines than at those of the Balmer lines, making Paschen-derived reddenings less dependent on the assumed extinction curve.

Prior to the launch of the \textit{James Webb Space Telescope} (\textit{JWST}), Paschen lines at $z\sim1-3$ were studied in only a few lensed sources \citep{Finkelstein2011, Papovich2011}, as they are intrinsically weaker than the Balmer lines and are redshifted out of the atmospheric transmission windows. With \textit{JWST}, NIRCam/grism and NIRSpec observations have enabled robust detections of Paschen lines in unlensed high-redshift galaxies \citep{Reddy2023, Neufeld2024, Seille2024, Reddy2025}. These studies highlighted the importance of accurately accounting for slit losses and relative flux calibration when studying high-redshift galaxies using multiple nebular emission lines, particularly when combining Balmer and Paschen lines which are widely separated in wavelength.

By combining new NIRCam/grism observations of Paschen lines with ground-based MOSFIRE measurements of Balmer lines in the same galaxies, we constructed a sample of galaxies with robust detections of both Balmer and Paschen lines. The NIRCam/grism observations were obtained from the First Reionization Epoch Spectroscopically Complete Observations survey \citep[FRESCO; PID 1895;][]{Oesch2023}. These observations used slitless spectroscopy and include emission maps that trace the spatial distribution of Paschen emission, thereby mapping the star-forming regions within the galaxies. In this paper, we use these data to constrain nebular reddening and SFRs using both Balmer and Paschen lines. This approach allows us to investigate the conditions under which the Balmer and Paschen lines yield significantly different inferences of reddening and SFRs. Additionally, the emission maps enable us to study the spatial distribution of star-forming regions within galaxies and to evaluate the accuracy of slit-loss corrections applied to the Balmer lines in previous studies.

The paper is organized as follows. In Section \ref{sec:data}, we describe the observations and sample selection. Section \ref{sec:Slit-Loss Corrections} examines morphological differences between Paschen emission and continuum emission observed in the F210M and F444W bands, and presents revised slit-loss corrections derived from Paschen emission maps. In Section \ref{sec:Nebular Reddening}, we compare nebular reddenings derived from the Balmer decrement and Paschen lines. Section \ref{sec:SFRs} compares SFRs derived from dust-corrected H$\alpha$, Pa$\alpha$, Pa$\beta$, and UV-based indicators. In Section \ref{sec:Reddening Correlations}, we examine how nebular reddening correlates with SED-inferred reddening, stellar mass, metallicity, UV SFR, and recombination line-derived SFR. In Section \ref{sec:Conclusion}, we summarize our findings. Throughout this work, we adopt a standard cosmology with $\Omega_\Lambda=0.7$, $\rm\Omega_M=0.3$, and $\rm H_0=70\ km\ s^{-1}\ Mpc^{-1}$.

\section{Data} \label{sec:data}

The spectroscopic data were drawn from two surveys: Balmer line measurements came from the MOSFIRE Deep Evolution Field (MOSDEF) survey \citep{Kriek2015}, while Paschen line data were taken from the FRESCO survey \citep{Oesch2023}. Previous studies using MOSDEF data constrained nebular reddening and SFRs in $z\sim2$ galaxies via the Balmer decrement \citep[e.g.,][]{Reddy2015, Shivaei2015, Shivaei2016, Reddy2020, Shivaei2020, Fetherolf2021, Lorenz2023, Reddy2023_2}. The addition of Paschen line measurements from FRESCO's \textit{JWST} grism observations expands the scope of nebular reddening and SFR analysis by enabling direct comparisons between estimates based on the commonly used Balmer decrement and those derived from decrements that include a Paschen line. Moreover, the FRESCO emission maps provide unprecedented spatially-resolved information on the Paschen emission, enabling an analysis of the spatial distribution of star-forming regions within galaxies.

\subsection{MOSDEF Survey} \label{subsec:mosdefsurvey}

Our analysis used data from the MOSDEF survey, which targeted $\sim$1500 \textit{H}-selected galaxies and active galactic nuclei (AGN) in the CANDELS fields \citep{Grogin2011, Koekemoer2011}. The survey obtained moderate-resolution ($R\sim3000-3650$) near-IR spectra using the MOSFIRE instrument on the 10\,m Keck I telescope over 48.5 nights. Targets were selected for observation based on external spectroscopic redshifts within three intervals: $z=1.37-1.70$, $2.09-2.61$, and $2.95-3.80$. These ranges were chosen such that bright rest-frame optical emission lines ($\rm[O\,\text{\small II}]$, H$\beta$, $\rm[O\,\text{\small III}]$, H$\alpha$, $\rm[N\,\text{\small II}]$, and $\rm[S\,\text{\small II}]$) fell within atmospheric transmission windows. The MOSDEF sample spans a wide range of stellar masses ($\sim\!10^9-10^{11.5}\ M_\odot$) and SFRs ($\sim\!10^0-10^3\ M_\odot\ \mathrm{yr^{-1}}$). For more details, we refer readers to \cite{Kriek2015}.

Following the procedure outlined in \cite{Reddy2015}, Balmer line fluxes were measured from the 1D spectra by fitting H$\beta$ with a Gaussian profile and the H$\alpha$ and $\rm[N\,\text{\small II}]$ doublet with three Gaussian profiles, all on top of a continuum determined from the best-fit spectral energy distribution (SED) to the photometry, as described in Section \ref{subsec:SEDFitting}. This method accounts for stellar absorption under the Balmer lines. To estimate uncertainties, each spectrum was perturbed 1000 times using its error spectrum, and fluxes were measured for each iteration. The final flux and its uncertainty were calculated as the average and standard deviation of these 1000 measurements, respectively.

Because the Balmer line fluxes were measured using slit-based observations, slit-loss corrections were required for the MOSDEF data to accurately constrain nebular reddening and SFRs. These corrections were derived using F160W postage stamps from the CANDELS imaging \citep{Skelton2014}, under the assumption that the line emission has a morphology similar to the F160W continuum emission. For further details on the slit-loss corrections, see \cite{Reddy2015}. These corrections are reevaluated in Section \ref{subsec:Slit Corrections} in the context of the nebular line emission maps provided by the FRESCO survey.

\subsection{FRESCO Survey} \label{subsec:frescosurvey}

Our analysis also included data from the FRESCO survey, which conducted \textit{JWST} NIRCam/grism observations between November 2022 and February 2023. The survey covered ${\sim60}\,\rm arcmin^2$ in the GOODS North and GOODS South fields using the F444W filter. FRESCO observed with grismR, covering a wavelength range of 3.8 to 5.0\,$\mu$m, with a spectroscopic exposure time of 7\,ks per pointing. The observations achieved a $5\sigma$ line sensitivity of $2.8\times10^{-18}\,\mathrm{erg\ s^{-1}\ cm^{-2}}$ at a resolution of $R\sim1600$ (for compact sources) and produced emission line maps with a spatial resolution of ${\sim}\,0.15$\arcsec and a pixel scale of 0.05\arcsec. The survey also obtained medium-band images with the F182M and F210M filters. For direct imaging, the exposure time was 0.9\,ks per pointing, reaching $5\sigma$ depths of 28.4, 28.2, and 28.2 mag for the F182M, F210M, and F444W filters, respectively (measured in 0.32\arcsec\ diameter circular apertures). For additional details on the survey, we refer readers to \cite{Oesch2023}. The survey data are publicly available in MAST: \hyperlink{https://archive.stsci.edu/doi/resolve/resolve.html?doi=10.17909/gdyc-7g80}{10.17909/gdyc-7g80}.

The data were reduced using the {\tt GRIZLI} software \citep{Brammer2019, Brammer2021}. Continuum emission was subtracted by applying a median filter with a 12-pixel central gap along each row \citep{Kashino2023}, which minimizes self-subtraction in the case of strong emission lines. The 1D spectra were then extracted using optimal extraction \citep{Horne1986} on the F444W+F210M segmentation maps. Paschen line fluxes (Pa$\alpha$ and Pa$\beta$) were measured using a bootstrap method similar to that used for H$\alpha$ and H$\beta$ fluxes in the MOSDEF data. Each science spectrum was perturbed by its error spectrum 1000 times, and fluxes were determined by fitting a Gaussian profile on top of a linear continuum in each perturbed iteration. The final flux and uncertainty were calculated as the median and standard deviation of the 1000 measurements, respectively.

To account for the minimal amount of stellar absorption under the Paschen lines, a Gaussian profile was fit to each galaxy's best-fit SED at the wavelength of each line to estimate the absorbed flux. The measured absorption was added to the observed line flux to derive the final Paschen flux. The stellar absorption was calculated separately because the FRESCO data lack the spectral resolution needed to simultaneously fit both the emission and absorption components, and because the continuum is typically not detected in the grism spectra. It is important to note that stellar absorption features may have broader intrinsic line widths than nebular emission lines, as atomic velocities in stellar atmospheres are typically much higher than the velocity dispersion within H\,$\rm\text{\small II}$ regions. This difference can lead to an overestimation of the absorption correction when using the above approach. However, when comparing the Gaussian fits to the science spectra with those to the SED modeling, we find that on average only 2\% of the absorbed flux lies farther than $3\sigma$ from the line center, where $\sigma$ is the standard deviation of the Gaussian fit to the emission line. Moreover, the absorption correction accounts for only 4\% of the final flux on average, suggesting that variations in intrinsic line widths have a negligible effect on the corrected flux measurements.

\subsection{Sample Selection} \label{subsec:sampleselection}

\begin{figure}

\includegraphics[width=0.45\textwidth]{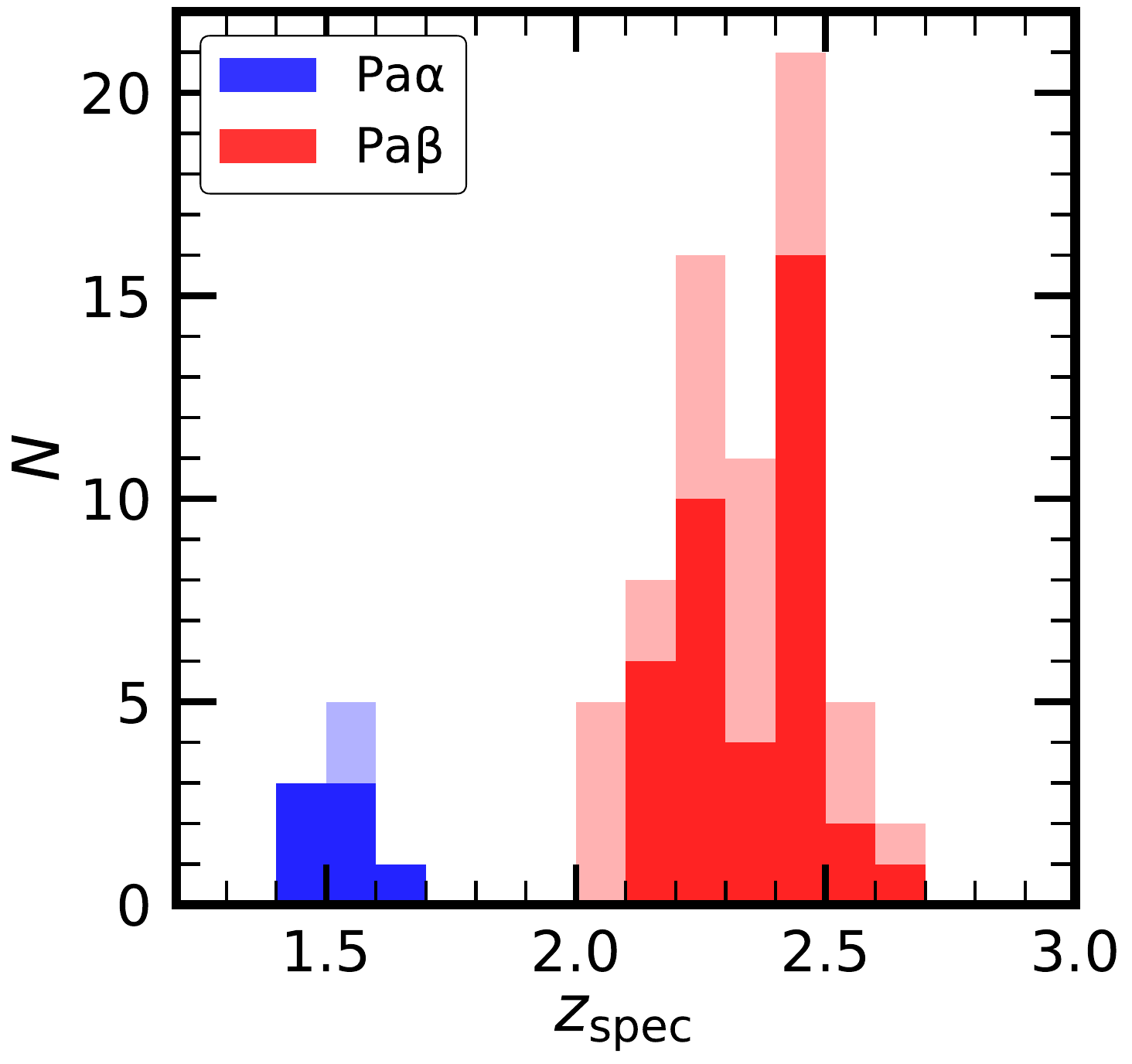}

\caption{Spectroscopic redshift histogram of the galaxy sample used in the analysis. The lighter blue and red bars indicate all galaxies with Pa$\alpha$ (9 objects) and Pa$\beta$ (68 objects) coverage, respectively. The darker blue and red bars represent galaxies with ${\geq}3\sigma$ detections of Pa$\alpha$ (7 objects) and Pa$\beta$ (39 objects), respectively.}
\label{fig: Redshift Histograms}

\end{figure}

Our analysis used galaxies observed in both the MOSDEF and FRESCO surveys. Each galaxy was required to have coverage of both the H$\alpha$ and H$\beta$ lines in the MOSDEF spectra, as well as either Pa$\alpha$ or Pa$\beta$ in the FRESCO data. Galaxies with significant skyline contamination of the Balmer lines or flagged as AGN (based on having $\rm[N\,\text{\small II}]/H\alpha\geq0.5$) were excluded \citep{Coil2015, Azadi2017}, resulting in a sample of 77 galaxies: 9 with Pa$\alpha$ coverage and 68 with Pa$\beta$ coverage. For the analysis of individual galaxies, no signal-to-noise (S/N) threshold was imposed on H$\alpha$ or H$\beta$, but we required Pa$\alpha$ and Pa$\beta$ detections with S/N $\geq 3$. This criterion yielded 7 galaxies with ${\geq}3\sigma$ Pa$\alpha$ detections and 39 galaxies with ${\geq}3\sigma$ Pa$\beta$ detections, comprising a final sample of 46 galaxies for individual analysis. All 46 of these galaxies have ${\geq}3\sigma$ H$\alpha$ detections and 38 have ${\geq}3\sigma$ H$\beta$ detections; all galaxies have ${\geq}2\sigma$ H$\beta$ detections. Within this individually-analyzed sample, many Paschen lines were detected well above the minimum S/N threshold, with 35 galaxies exhibiting Paschen line detections with S/N $\geq5$. All galaxies, regardless of the S/N of their lines, were included in the construction of composite spectra, as described in Section \ref{subsec:Compositespectra}. The full sample of 77 galaxies spans a redshift range of $1.43\leq z\leq1.65$ for the Pa$\alpha$-detected galaxies and $2.05\leq z\leq2.65$ for the Pa$\beta$-detected galaxies. The redshift distribution of the sample is shown in Figure \ref{fig: Redshift Histograms}. The smaller number of Pa$\alpha$-detected galaxies reflects the limited redshift range where H$\alpha$ and H$\beta$ were covered in the MOSDEF observations and Pa$\alpha$ was simultaneously covered in the FRESCO observations. Specifically, the FRESCO observations had coverage of Pa$\alpha$ from $z\sim1.02-1.66$, while the MOSDEF survey targeted no galaxies with redshifts below $z=1.37$, as this is the lowest redshift at which H$\beta$ could be observed in the \textit{J}-band filter.

\subsection{SED Fitting} \label{subsec:SEDFitting}

Reddening of the stellar continuum, stellar mass, and UV SFR were estimated using SED modeling of multi-wavelength photometric data from 3D-HST catalogs \citep{Skelton2014, Momcheva2016}. The modeling employed the Binary Population and Stellar Synthesis (BPASS) v2.2.1 model \citep{Eldridge2017, Stanway2018}, assuming the SMC extinction curve, constant star-formation histories, and a \cite{Chabrier2003} initial mass function (IMF). A lower age limit of 50 Myr was adopted, based on the typical dynamical timescale of $z\sim2$ galaxies \citep{Reddy2012}. Stellar population parameters and their uncertainties were calculated by perturbing the photometric data 100 times, refitting the perturbed photometry, and deriving the median and standard deviation of the resulting distributions. For further details on the modeling approach, see \cite{Reddy2015}. SED fitting for the sample yielded stellar masses ranging from $\sim\!10^{8.8}-10^{11.0}\ M_\odot$ and SFRs ranging from $\sim\!10^{0.3}-10^{1.6}\ M_\odot\ \mathrm{yr^{-1}}$.

\subsection{Composite Spectra} \label{subsec:Compositespectra}

To account for galaxies without significant Paschen line detections, high S/N composite spectra were constructed by combining individual MOSDEF and FRESCO observations. Each spectrum was shifted to the rest frame using the spectroscopic redshift, converted to luminosity density, and interpolated onto a common wavelength grid. At each wavelength point, the composite luminosity density and its uncertainty were calculated as the mean and standard deviation of all resampled spectra after applying $3\sigma$ clipping. Integrated line luminosities for the composite spectra were measured using the same Gaussian fitting method applied to individual MOSDEF and FRESCO spectra. When calculating line flux ratios from the composite spectra, each galaxy's spectrum was normalized by the flux of the line appearing in the denominator of the ratio (e.g., H$\beta$ for the Balmer decrement or H$\alpha$ for the Paschen-to-H$\alpha$ ratios) before combining the spectra. To account for uncertainties, 1000 realizations of the composite spectrum were created. In each iteration, the science spectra were perturbed by their error spectra, and galaxies were randomly selected with replacement to account for sample variance. Line luminosities were measured for each realization, and the final composite luminosity and uncertainty were taken as the median and standard deviation of the measured luminosities across all iterations, respectively. Composite spectra were constructed using all galaxies in the sample, including those with formally undetected Paschen lines.

\section{Emission Maps}\label{sec:Slit-Loss Corrections}

The F444W+F210M continuum maps from the FRESCO survey trace the spatial distribution of rest-frame optical and near-IR continuum emission. At these wavelengths, the continuum is dominated by older, less massive stars in the typical galaxy within the sample. In contrast, the Paschen emission originates from the youngest, most massive stars, providing a tracer of recent star formation. Together, these emission maps enable a direct comparison between the spatial distributions of newly formed stars and of the older stellar population. The maps also allow us to evaluate how discrepancies between the two distributions affect slit-loss corrections that assume the continuum and emission-line morphologies are identical.

\subsection{Paschen and Continuum Spatial Distributions} \label{subsec:SpatialDistribution}

We used the Paschen and F444W+F210M continuum maps to measure the spatial extent of the nebular emission relative to the continuum \citep[e.g.,][]{Nelson2016, Matharu2022, Matharu2024, Neufeld2024}. Half-light areas ($\rm A_e$) were calculated from each map by summing the flux pixel by pixel in order of decreasing flux until half of the light was included. The number of pixels in the sum was converted to $\rm arcsec^2$ using the pixel scale, and then to $\rm kpc^2$ based on the spectroscopic redshift. Half-light area uncertainties were estimated by perturbing each emission map 1000 times, measuring the half-light area for each iteration, and adopting the standard deviation of the resulting distribution as the uncertainty. Because the surface brightness of the Paschen emission may be falling off rapidly, half-light areas were also measured from three composite maps. These composites were constructed by sorting the full sample of 77 galaxies into three equal-sized bins based on Paschen half-light area, and then stacking the emission maps within each bin. The composite maps enable us to account for extended emission that may be missed when examining individual galaxies.

\begin{figure}

\includegraphics[width=0.45\textwidth]{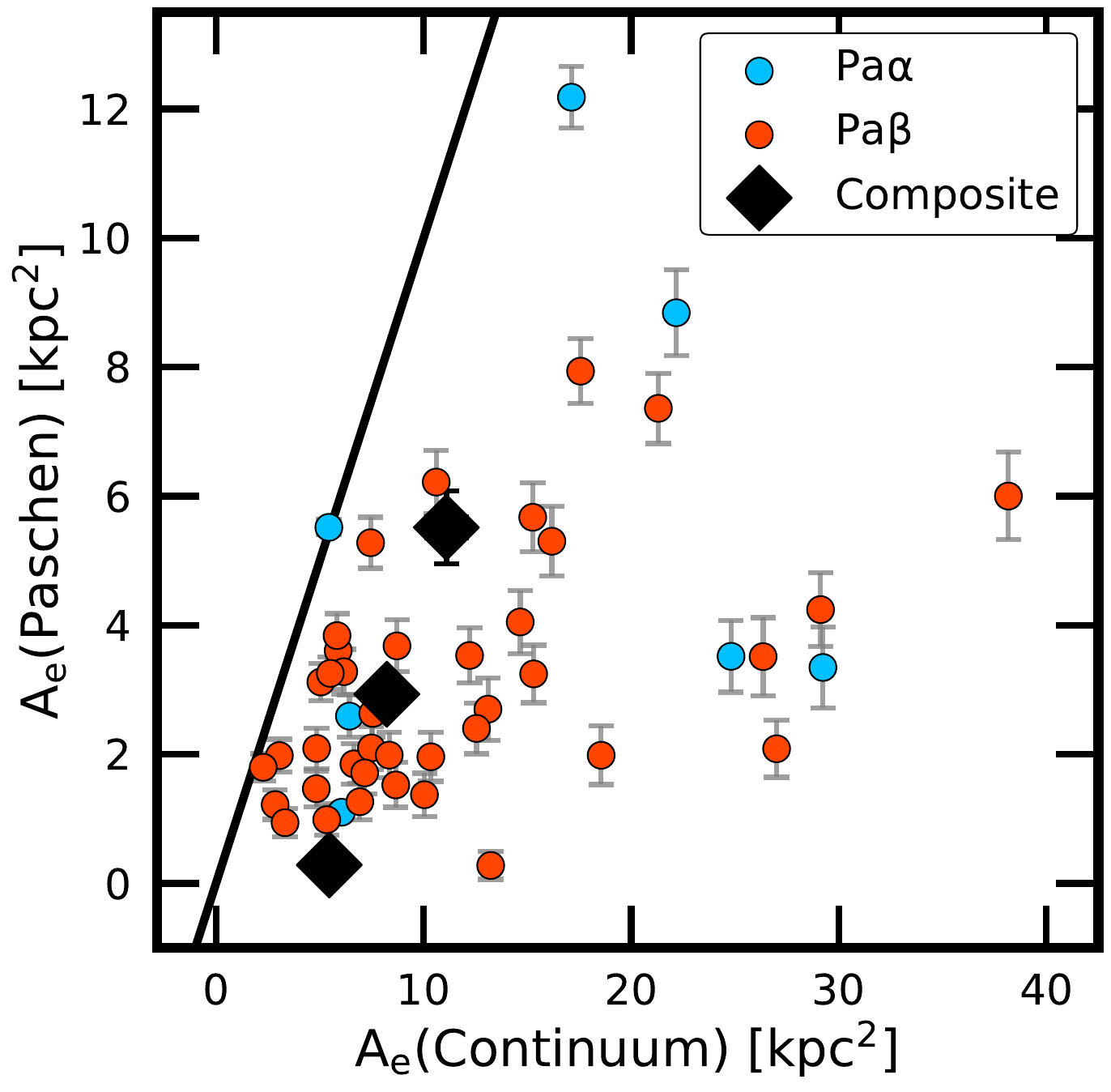}

\caption{Half-light areas derived from the Paschen and F444W+F210M continuum maps. The solid line indicates equality between the two measurements. Diamonds denote composite values obtained after sorting the full sample of 77 galaxies into three equal-sized bins based on Paschen half-light area. Paschen emission exhibits smaller areas than the continuum emission in all but one galaxy, with the composites likewise showing smaller Paschen emission areas.}
\label{fig: Half-Light Areas}

\end{figure}

The resulting half-light areas are shown in Figure \ref{fig: Half-Light Areas}. All but one galaxy in the sample exhibit more concentrated Paschen emission morphologies compared to the continuum, with 19 galaxies showing Paschen half-light areas less than 25\% of their continuum half-light areas. The single galaxy with a more concentrated continuum morphology has a Paschen half-light area that is only 2\% larger than that of the continuum. All three composites also exhibit more concentrated Paschen emission, with continuum half-light areas that are 19.2, 2.8, and 2.0 times larger than the corresponding Paschen half-light areas. This suggests that the trend of more concentrated Paschen emission morphologies persists even when accounting for faint, extended emission that may be missed in individual galaxies. The smaller Paschen half-light areas relative to the continuum indicate that current star formation is more spatially concentrated than the older stellar population traced by the continuum.

These results are particularly relevant in light of previous studies showing that the effective radius of H$\alpha$ emission is typically greater than or equal to that of the stellar continuum when measured using \texttt{GALFIT} \citep{Peng2002, Peng2010}, indicative of inside-out star formation \citep{Nelson2016, Matharu2022, Matharu2024}. \texttt{GALFIT} yields a larger Balmer effective radius when a significant fraction of the observed emission originates from star-forming regions located far from the center of the fit to the galaxy. In contrast, the method used here to calculate half-light area does not account for the distance of each pixel from the light-weighted center of the galaxy, making it insensitive to the radial distribution of nebular emission. To investigate the radial distribution of the Paschen emission, we used the normalized second-order moment of the brightest 20\% of the emission, known as the $M_{20}$ parameter \citep{Lotz2004}. The total second-order moment is defined as:

\begin{equation}
M_{\rm tot}=\sum\limits^{n}_{i}{M_i}=\sum\limits^{n}_{i}{f_i[(x_i-x_c)^2+(y_i-y_c)^2]},
\label{eq:M_tot}
\end{equation}

\noindent where $f_i$ is the flux in each pixel, ($x_i$, $y_i$) is the location of each pixel, and ($x_{c}$, $y_{c}$) is the galaxy's center, which is defined as the position that minimizes $M_{\rm tot}$. The $M_{20}$ parameter is calculated by ordering the pixels from highest to lowest flux, summing $M_i$ over the brightest pixels until the cumulative flux reaches 20\% of the total galaxy flux, and then normalizing by $M_{\rm tot}$:

\begin{equation}
M_{20}=\log{\left(\frac{\sum\limits{M_i}}{M_{\rm tot}} \right)},\ \textrm{while}\sum\limits_{i}{f_i}<0.2f_{\rm tot},
\label{eq:M_20}
\end{equation}

\noindent where $f_{\rm tot}$ is the total flux of the galaxy's pixels. Normalizing by $M_{\rm tot}$ removes dependence on the galaxy's total flux and size. To estimate uncertainties, each emission map was perturbed 1000 times, and the $M_{20}$ value was measured in each iteration. The final $M_{20}$ value and its uncertainty were calculated as the mean and standard deviation of these 1000 measurements, respectively. $M_{20}$ values are typically negative, with lower (more negative) values indicating that the brightest pixels are concentrated near the galaxy's center, and higher (less negative) values indicating that the brightest pixels are farther from the center. The $M_{20}$ parameter provides a quantitative measure of how the brightest regions are spatially distributed relative to the center of the galaxy.

\begin{figure*}

\includegraphics[width=0.95\textwidth]{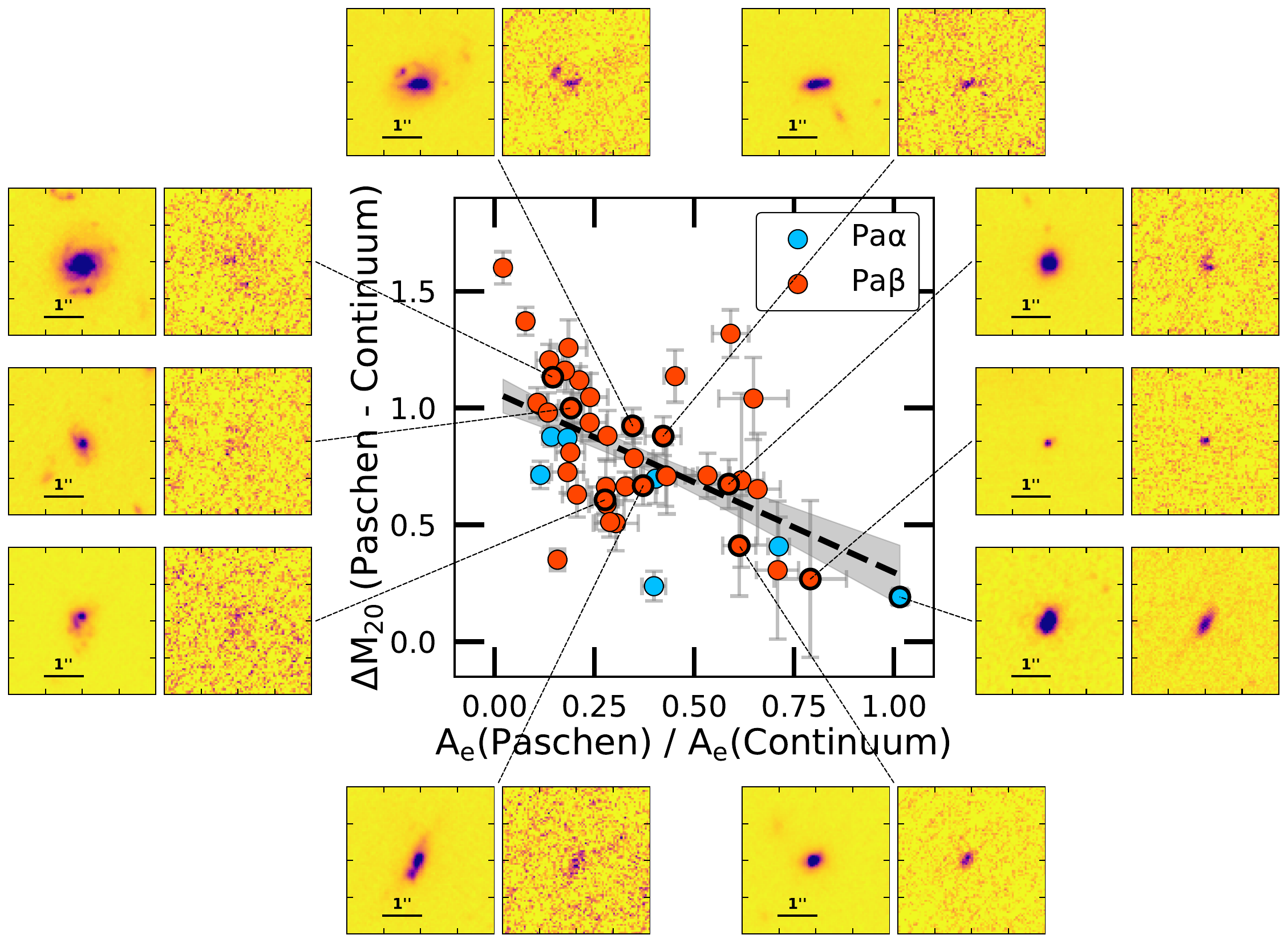}

\caption{Differences between Paschen $M_{20}$ and continuum $M_{20}$ values plotted against the ratios of Paschen to continuum half-light areas. The dashed line represents the best-fit relation, with the shaded region showing the 68\% confidence interval. Emission maps are shown for a selection of galaxies in the sample. For each galaxy, the left panel shows the continuum emission constructed from a stack of the F444W and F210M images, while the right panel shows the Paschen emission line map. Galaxies with more concentrated Paschen emission areas relative to the continuum tend to have Paschen emission located farther from the galaxy's center compared to the continuum.}
\label{fig: M20 and Half-Light Areas}

\end{figure*}

We calculated $M_{20}$ values from both the Paschen and continuum emission maps. Figure \ref{fig: M20 and Half-Light Areas} shows the difference between these measurements plotted against the ratio of Paschen half-light area to continuum half-light area, along with emission maps representative of typical Paschen emission and continuum morphologies across the sample. For all galaxies in the sample, Paschen $M_{20}$ values are greater than continuum $M_{20}$ values, indicating that the brightest regions of Paschen emission are more distant from the galaxy's center than the brightest continuum regions. This is indicative of inside-out star formation, as the youngest stars traced by the Paschen emission tend to be located farther from the central region of the galaxy than older stellar populations traced by the continuum. A Spearman test between the difference in Paschen and continuum $M_{20}$ values and the ratio of Paschen to continuum half-light areas yields a p-value of $2.70\times10^{-4}$ and a correlation coefficient of $-0.51$, indicating a negative correlation with $3.6\sigma$ significance. This result indicates that galaxies with Paschen emission that is more spatially concentrated than the continuum tend to have Paschen emission that is located farther away from the galaxy's center. This is consistent with young stars forming preferentially toward the outskirts of galaxies, as expected in inside-out growth scenarios. The significant presence of nebular emission in the outskirts of galaxies is consistent with the larger effective radii of Balmer emission reported in previous studies.

\subsection{Slit-Loss Corrections} \label{subsec:Slit Corrections}

Since the Paschen lines were observed in the grism mode, they do not suffer from slit losses. In contrast, the Balmer lines were measured with ground-based slit spectroscopy and, as discussed in \cite{Reddy2015}, their fluxes were corrected for slit loss assuming that the line-emission morphology is similar to that of the F160W continuum. If the nebular emission is more spatially concentrated than the continuum, the slit-loss correction may overestimate the fraction of Balmer emission outside the slit. This will lead to overestimated Balmer fluxes and consequently a smaller Paschen-to-Balmer line ratio. As a result, galaxies with more concentrated nebular emission can exhibit Paschen-derived reddening values lower than their intrinsic reddening, due to inaccurate slit-loss corrections. The Balmer decrement will also be affected by inaccuracies in slit-loss corrections, since H$\alpha$ and H$\beta$ lie in different ground-based filters, and therefore were observed at separate times and under different seeing/weather conditions. However, the slit-loss corrections for H$\alpha$ and H$\beta$ used the same assumptions regarding the emission morphology, so any overestimation tends to cancel out in the ratio. Additionally, the slit-loss correction factors for H$\alpha$ and H$\beta$ differ by no more than 5\% across the sample, indicating that the Balmer decrement is minimally affected by inaccurate slit-loss corrections.

\begin{figure*}

\includegraphics[width=0.95\textwidth]{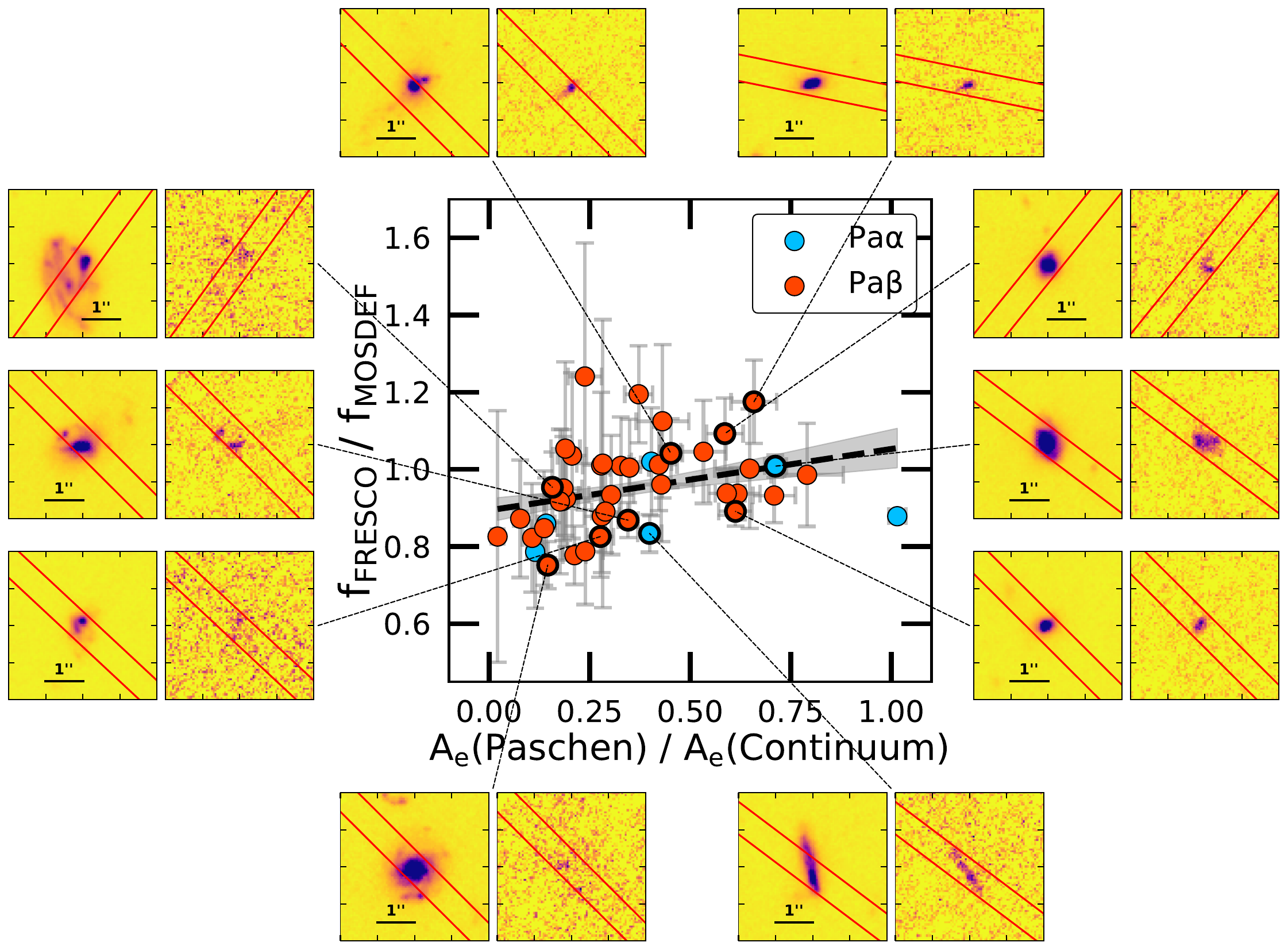}

\caption{Ratios of revised slit-loss correction factors to those from the MOSDEF survey, plotted against the ratios of Paschen to continuum half-light areas. The dashed line represents the best-fit relation, with the shaded region showing the 68\% confidence interval. Emission maps are shown for a selection of galaxies in the sample. For each galaxy, the left panel shows the continuum emission constructed from a stack of the F444W and F210M images, while the right panel shows the Paschen emission line map. Red lines mark the position of the slit during the MOSDEF observations, and 1\arcsec\ scaling is shown in each continuum map. Galaxies with more concentrated Paschen emission areas relative to the continuum tend to have MOSDEF slit-loss corrections that overestimated the amount of excluded Balmer emission; for these galaxies, the slit contains most of the Paschen emission but excludes a significant fraction of the continuum.}
\label{fig: Slit Fluxes and Half-Light Areas}

\end{figure*}

We used the Paschen and F444W+F210M continuum maps to investigate correlations between relative emission sizes and overestimation of slit-loss corrections. To determine the ``true'' slit-loss corrections, the total Paschen flux for each galaxy was first calculated using the Paschen emission maps. To match the MOSDEF seeing conditions, each map was convolved with a Gaussian kernel to reproduce the spatial resolution of the seeing-limited MOSDEF observations. We then placed each map in the MOSDEF slit frame, measured the flux within the MOSFIRE slit aperture, and defined the correction factor ($\rm f_{FRESCO}$) as the ratio of the total flux to the flux contained in the slit. Figure \ref{fig: Slit Fluxes and Half-Light Areas} shows the ratios of the revised slit-loss correction factors to those used in the MOSDEF survey ($\rm f_{MOSDEF}$), plotted against the ratios of Paschen half-light areas to continuum half-light areas. Figure \ref{fig: Slit Fluxes and Half-Light Areas} also shows the positions of the slits in the MOSDEF observations overlaid on emission maps representative of typical Paschen emission and continuum morphologies across the sample. For clarity, we excluded one galaxy where the slit was positioned such that most of the Paschen emission lay outside the slit, resulting in an outlier $\rm f_{FRESCO}/f_{MOSDEF}$ value of 1.75. For the remaining galaxies, the $\rm f_{FRESCO}/f_{MOSDEF}$ values range from 0.75 to 1.24, with an average value of 0.95. MOSDEF slit-loss corrections are generally valid on average. However, a Spearman test yields a p-value of $1.24\times10^{-3}$ and a correlation coefficient of $0.46$, corresponding to a positive correlation with $3.2\sigma$ significance. This correlation indicates that galaxies with more concentrated Paschen emission relative to the continuum tend to have overestimated MOSDEF slit-loss corrections. These overestimations bias the Balmer fluxes in the sample toward higher values. The original MOSDEF slit-loss corrections were replaced with the revised corrections for the H$\alpha$ and H$\beta$ fluxes, and these updated fluxes were used for the remainder of the analysis.

It is important to note that the new slit-loss corrections assume that the Balmer emission morphology matches that of the Paschen emission. This assumption is more accurate than the MOSDEF slit-loss corrections, which assume the Balmer emission follows the continuum morphology. However, inaccuracies can still arise in galaxies where a significant fraction of star formation is optically thick to the Balmer lines and thus only visible in the Paschen lines. In such cases, the observed Balmer emission is limited to less dusty lines of sight and would nominally have a smaller spatial extent relative to the Paschen emission. A more spatially concentrated Balmer morphology relative to the Paschen morphology would cause the new slit-loss corrections to overestimate the fraction of Balmer emission outside the slit, leading to an overestimation of the Balmer flux, similar to the bias seen in the MOSDEF corrections. Along these lines, \cite{Lorenz2025} found that slight offsets between H$\alpha$ and Pa$\beta$ emission ($<0.1$\arcsec) are common at $z\sim2$, particularly in the dustiest galaxies with high nebular reddening. However, the differences between Balmer and Paschen morphologies are smaller than those between Balmer and continuum morphologies. For this reason, the new slit-loss corrections more accurately reflect the Balmer morphologies than those used in the MOSDEF survey. Future studies comparing Balmer and Paschen morphologies can further explore how differences between the two morphologies affect inaccuracies in slit-loss corrections.

\section{Nebular Reddening} \label{sec:Nebular Reddening}

The ratios of hydrogen recombination lines are sensitive to dust reddening along the lines of sight toward H\,$\rm\text{\small II}$ regions, since wavelength-dependent attenuation causes the observed ratio to deviate from its intrinsic dust-free value. However, the Balmer decrement may be insensitive to reddening toward the most heavily dust-obscured regions that are optically thick in the Balmer lines. Because Pa$\alpha$ and Pa$\beta$ are less attenuated by dust than the Balmer lines, decrements including these Paschen lines are more sensitive to these heavily dust-obscured regions. Comparing nebular reddenings derived from the Balmer decrement with those from decrements including Pa$\alpha$ or Pa$\beta$ can therefore reveal the presence of dusty regions missed by the Balmer decrement. As in most previous studies, we assumed that dust attenuates light uniformly along all lines of sight. However, we also investigated how non-unity dust covering fractions affect the observed reddening toward nebular regions \citep[e.g.,][]{Reddy2025}.

\subsection{Observed Nebular Reddening}\label{subsec:Observed Nebular Reddening}

\begin{figure*}

\includegraphics[width=0.95\textwidth]{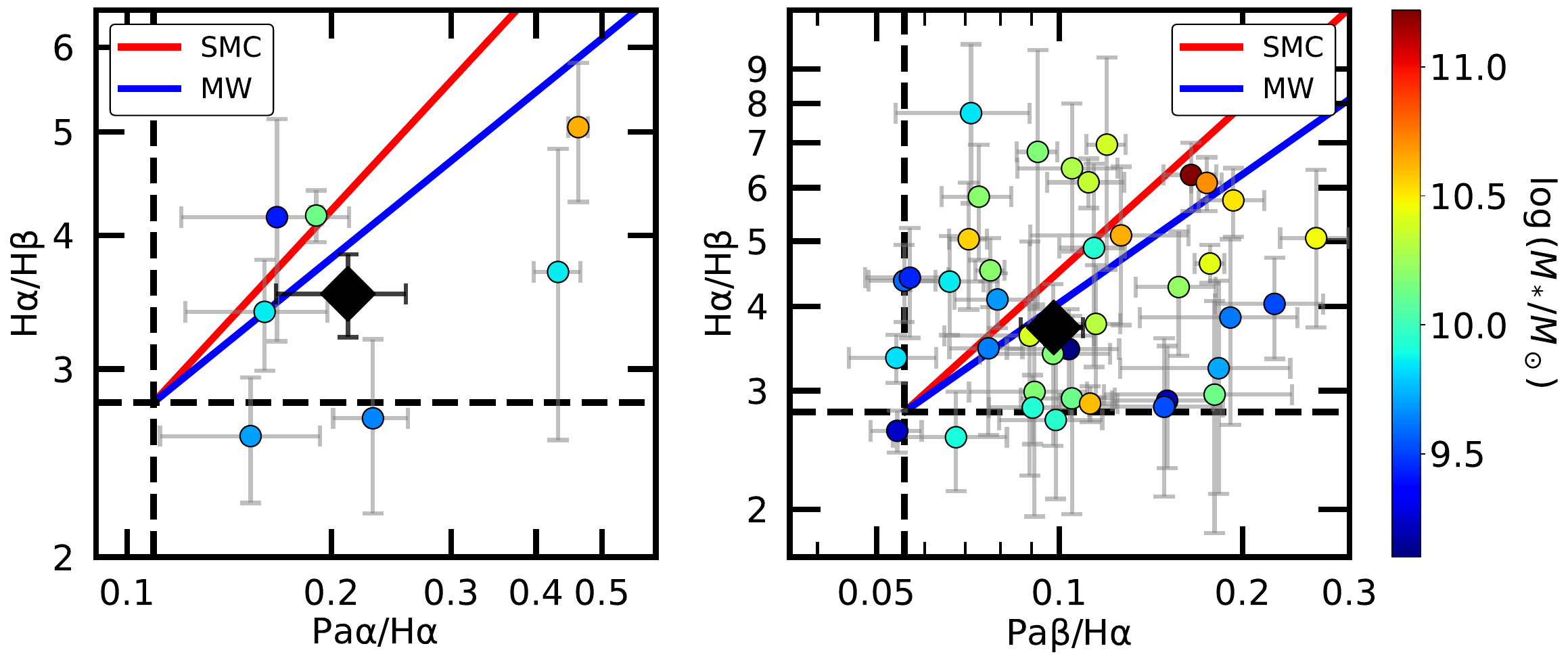}

\caption{Balmer decrement compared to Pa$\alpha$/H$\alpha$ (left) and Pa$\beta$/H$\alpha$ (right), color-coded by SED-derived stellar mass. Large diamonds indicate values measured from composite spectra of all galaxies with the required line coverage. The red and blue lines show predictions when assuming the SMC and MW extinction curves, respectively, while dashed lines indicate line ratios in the absence of dust attenuation.}
\label{fig: Recombination Ratios}

\end{figure*}

The nebular reddening for each galaxy was calculated using the Balmer decrement and the ratio of the detected Paschen line to $\rm H \alpha$, assuming the MW extinction curve \citep{Cardelli1989}. Unattenuated line ratios were calculated assuming Case B recombination, with $T=10,000$\,K and $n_e=100\ \rm cm^{-3}$ (see Table \ref{tab:ratios}). Figure \ref{fig: Recombination Ratios} shows the flux ratios used to derive nebular reddening, along with the ratios predicted when assuming the MW and SMC \citep{Gordon2003} extinction curves. Overall, the ratios for individual galaxies align more closely with the prediction of the MW curve, although all but four galaxies are consistent with both curves within $3\sigma$. Composite ratios fall within one standard deviation of the predicted values of both extinction curves. This is not surprising, as the MW and SMC curves have similar shapes at the wavelengths of the Balmer and Paschen lines. Employing multiple sets of recombination line ratios can provide additional constraints on the shape of the attenuation curve, as demonstrated in \cite{Reddy2020}, \cite{Rezaee2021}, and \cite{Reddy2025}.

\begin{table}\
\begin{center}

\caption{Hydrogen Recombination Lines and Relative Line Intensities}

\begin{tabular}{c|cc} 
\hline \hline

Line & $\lambda(\text{\AA})$ \footnote{Rest-frame vacuum wavelengths, from the Atomic Spectra Database website of the National Institute of Science and Technology (NIST) \url{https://www.nist.gov/pml/atomic-spectra-database}.}  & I \footnote{Intensity of line relative to $\rm H \alpha$ for Case B recombination. $T_e=10,000$\,K and $n_e=100\ \rm cm^{-3}$. Based on photoionization modeling with CLOUDY version 17.02 \citep{Ferland2017, Reddy2022}.} \\

\hline 
$\rm H \beta$ & 4862.71 & 0.358\\
$\rm H \alpha$ & 6564.60 & 1.000\\
$\rm Pa \beta$ & 12821.6 & 0.056\\
$\rm Pa \alpha$ & 18756.4 & 0.109\\
\hline \hline  
\end{tabular}%

\label{tab:ratios}
\end{center}
\end{table}

\begin{figure}

\includegraphics[width=0.45\textwidth]{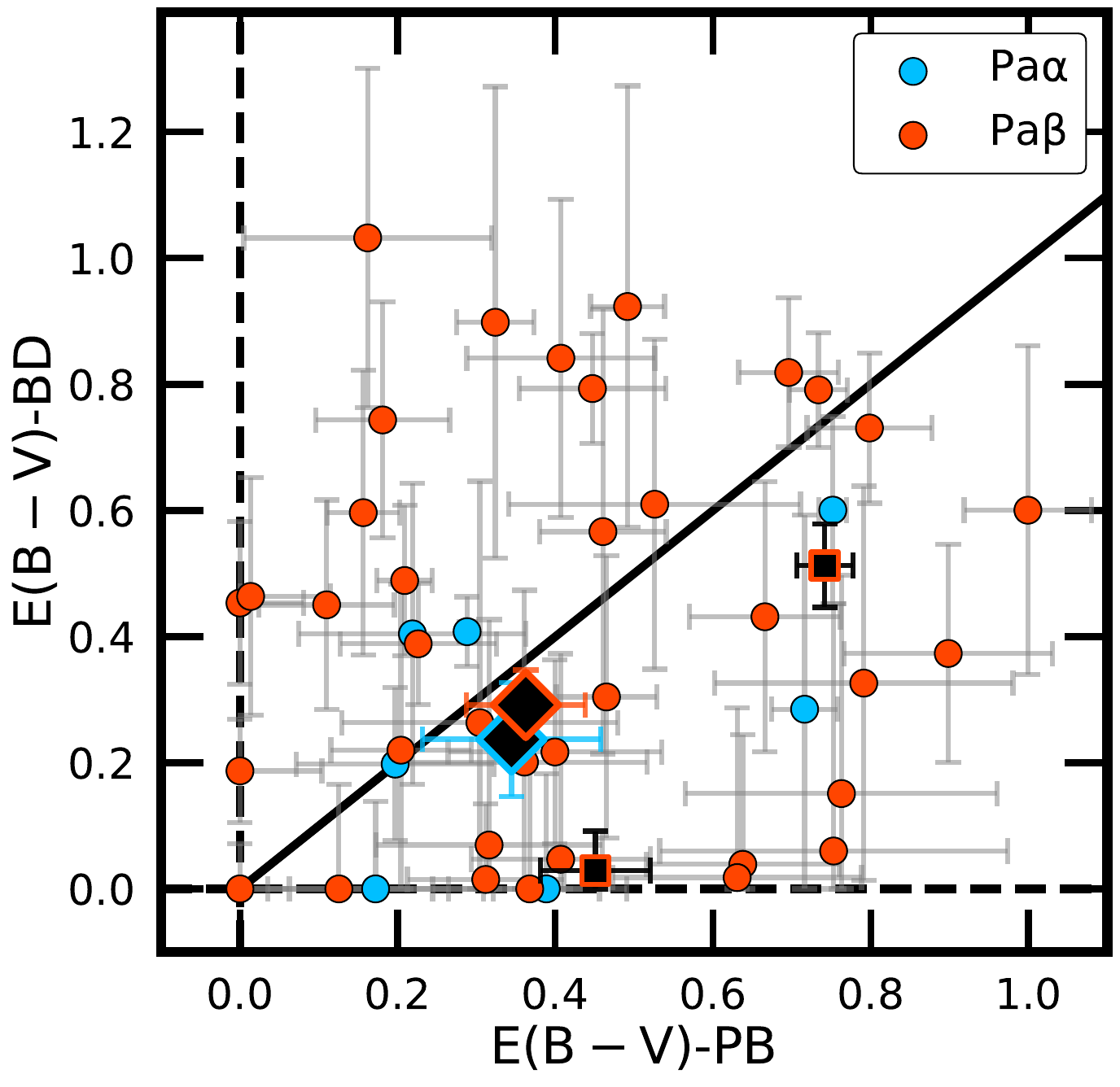}

\caption{Nebular reddening values derived from the hydrogen recombination line flux ratios shown in Figure \ref{fig: Recombination Ratios}, assuming the MW extinction curve. Blue and red points represent galaxies in the $\rm Pa \alpha$- and $\rm Pa \beta$-detected samples, respectively. Galaxies with one reddening value significantly larger than the other (${\geq} 3\sigma$) are marked with black squares. Large diamonds denote values from composite spectra. The dashed lines indicate zero attenuation, and the solid line marks equality between the two reddening estimates.}
\label{fig: Nebular Reddening; New Slit-Loss Corrections}

\end{figure}

The attenuation curve is defined as:

\begin{equation}
k_\lambda\equiv\frac{A_\lambda}{E(B-V)},
\label{eq:AttenuationCurve}
\end{equation}

\noindent where $A_\lambda$ is the total magnitudes of attenuation at wavelength $\lambda$ and $E(B-V)$ is the nebular reddening. Using this definition, reddening values were derived from the flux ratios following the methodology of \cite{Reddy2015}:

\begin{equation}
E(B-V)=\frac{2.5}{k_2-k_1}\log{\biggl(\frac{F_1/F_2}{F_{1,0}/F_{2,0}}\biggr)},
\label{eq:Reddening}
\end{equation} 

\noindent where $F_1/F_2$ is the observed flux ratio, $F_{1,0}/F_{2,0}$ is the intrinsic dust-free flux ratio, and $k_1$ and $k_2$ are the attenuation curve values at the wavelengths of the respective lines. For the remainder of this paper, $E(B-V)\textrm{-BD}$ refers to nebular reddening derived from the Balmer decrement, and $E(B-V)\textrm{-PB}$ refers to reddening derived from the ratio of Pa$\alpha$ or Pa$\beta$ to H$\alpha$. All observed flux ratios falling below their dust-free values are within one standard deviation of the dust-free value; these galaxies were assigned a reddening of zero. The calculated nebular reddening values are shown in Figure \ref{fig: Nebular Reddening; New Slit-Loss Corrections}. For individual galaxies, all but two have $E(B-V)\textrm{-PB}$ values consistent with their $E(B-V)\textrm{-BD}$ values within $3\sigma$. The other two galaxies show significantly larger $E(B-V)\textrm{-PB}$ values. Composite $E(B-V)\textrm{-PB}$ values are 45\% larger than $E(B-V)\textrm{-BD}$ for the Pa$\alpha$-detected sample and 24\% larger for the Pa$\beta$-detected sample; all composite $E(B-V)\textrm{-PB}$ values are within one standard deviation of the composite $E(B-V)\textrm{-BD}$ values. While there is substantial scatter between the two reddening estimates, most of this scatter can be attributed to measurement errors, as nearly all galaxies have consistent values within $3\sigma$. Measurement uncertainties are more significant for ratios involving the Paschen lines, since these lines are weaker than the Balmer lines.

Both the individual and composite reddening values show that $E(B-V)\textrm{-BD}$ and $E(B-V)\textrm{-PB}$ agree on average within the measurement uncertainties, indicating that the Balmer- and Paschen-derived reddenings generally trace dust attenuation toward the same nebular regions. However, the measurement uncertainties still allow for a small systematic offset between $E(B-V)\textrm{-BD}$ and $E(B-V)\textrm{-PB}$, as reported in previous studies \citep{Gimenez2022, Prescott2022, Reddy2023, Lorenz2025, Reddy2025}. Because both reddenings were determined using H$\alpha$, it is difficult to distinguish such an offset. Systematic deviations between $E(B-V)\textrm{-BD}$ and $E(B-V)\textrm{-PB}$ could be identified more clearly by using multiple measurements of Paschen lines to determine $E(B-V)\textrm{-PB}$ \citep[e.g.,][]{Reddy2023, Reddy2025}. Notably, two galaxies in the sample exhibit significantly higher $E(B-V)\textrm{-PB}$ values, reflecting the presence of heavily dust-enshrouded regions where star formation is opaque in the Balmer lines but still visible in the Paschen lines. Additionally, as discussed in Section \ref{subsec:Covering Fractions}, these two galaxies may contain patchy dust distributions that bias the observed Balmer decrement toward less reddened OB associations.

\subsection{Non-Unity Dust Covering Fractions}\label{subsec:Covering Fractions}

Several pieces of evidence support the presence of non-unity dust covering fractions toward OB associations in high-redshift star-forming galaxies. These include the detections of massive O-star P-Cygni photospheric lines (e.g., $\rm C\,\text{\small IV}\,\lambda\lambda1548,1550$ for $\gtrsim30\,M_\odot$ O stars and $\rm Si\,\text{\small IV}\,\lambda\lambda1393,1402$ in O supergiants) and Wolf-Rayet features (e.g., He\,$\rm\text{\small II}\,\lambda 1640$) in far-UV spectra \citep[e.g.,][]{Pettini2000, Shapley2003}, spatial offsets between UV and dust emission \citep[e.g.,][]{Willott2015, Pentericci2016, Schouws2022, Bowler2022, Inami2022}, and elevated Paschen-derived reddenings relative to Balmer decrement-derived reddenings \citep[e.g.,][]{Prescott2022, Reddy2023, Reddy2025}. Additionally, \cite{Lorenz2025} suggested that patchy dust geometries may explain $\lesssim0.1$\arcsec offsets observed between H$\alpha$ and Pa$\beta$ emission maps in galaxies at $z\sim2$.

\begin{figure*}

\includegraphics[width=0.95\textwidth]{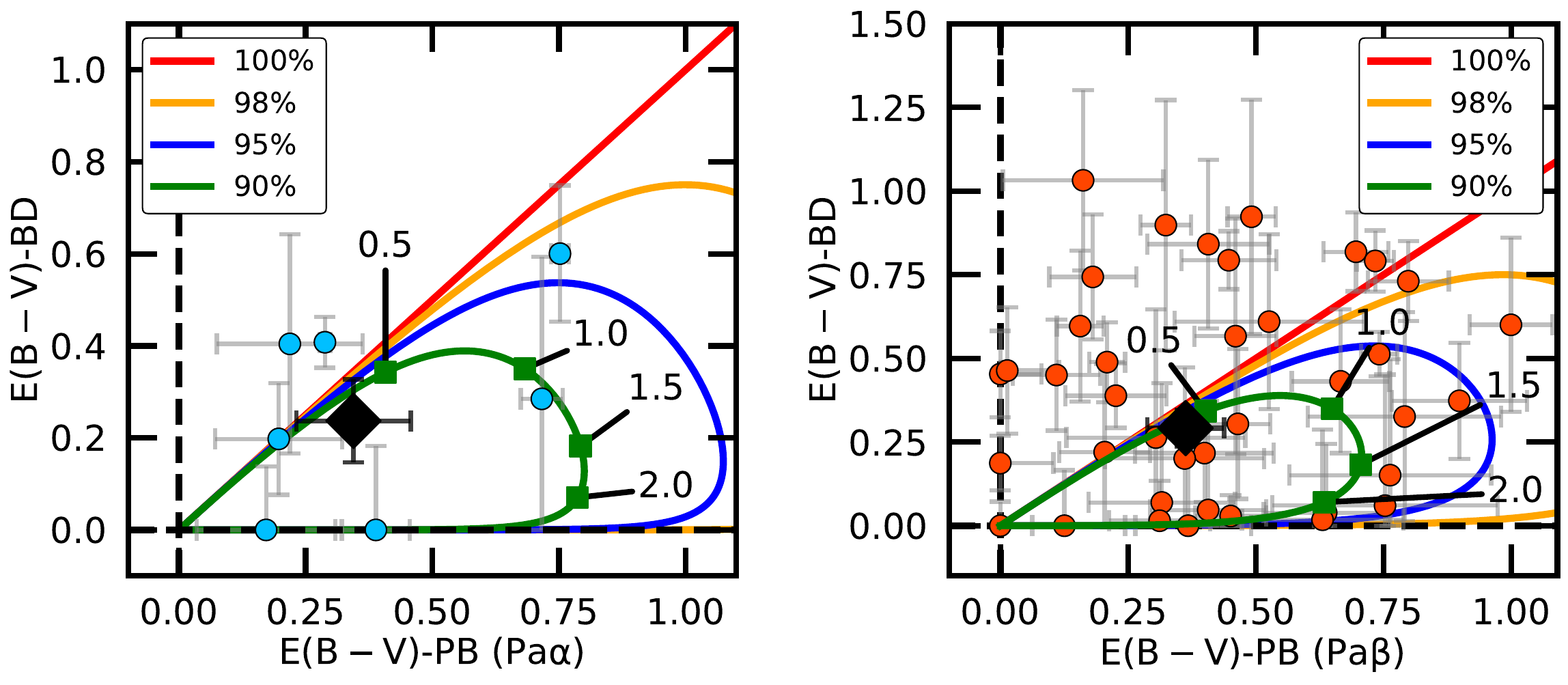}

\caption{Observed nebular reddening values for 100\%, 98\%, 95\%, and 90\% covering fractions, assuming the MW extinction curve, compared to the measured reddenings for the sample. Large diamonds indicate values from composite spectra. Dashed vertical and horizontal lines denote zero reddening. $E(B-V)$ values from the dust-obscured regions in the model are labeled along the 90\% covering fraction curve in intervals of $0.5$, up to a value of 2. As reddening toward the dust-covered regions increases, unattenuated emission increasingly dominates the observed flux, causing the measured reddening to decrease. This effect is more significant for $E(B-V)\textrm{-BD}$ than for $E(B-V)\textrm{-PB}$.}
\label{fig: Covering Fractions}

\end{figure*}

To explore the impact of non-uniform dust covering fractions on measured nebular reddening, we considered a scenario in which a fraction of the emission is unaffected by dust while the remainder is fully reddened. We used a methodology similar to that adopted by \cite{Reddy2016} to model the effect of a non-unity covering fraction of gas (and dust) on the stellar continuum, by applying the following equation:

\begin{equation}
F_{\rm\lambda,obs}=(1-f_{\rm cov})F_\lambda+f_{\rm cov}F_\lambda10^{-0.4A_\lambda},
\label{eq:covering_frac}
\end{equation}

\noindent where $F_{\rm\lambda,obs}$ is the observed flux density, $f_{\rm cov}$ is the dust covering fraction, $F_\lambda$ is the intrinsic flux density without dust attenuation, and $A_\lambda$ is the attenuation in magnitudes at wavelength $\lambda$ \citep{Reddy2015, Reddy2016, Prescott2022}. Figure \ref{fig: Covering Fractions} shows the predicted nebular reddening values for covering fractions of 100\%, 98\%, 95\%, and 90\%. For covering fractions below 100\%, greater attenuation toward the dust-obscured regions causes the observed flux to become increasingly weighted toward the unattenuated emission, biasing the measured reddening to lower values. When reddening toward the dust-obscured regions is sufficiently high, the unattenuated component dominates the total flux, and the observed reddening decreases toward zero as attenuation increases. The Paschen-based decrements are less biased toward the unattenuated emission than the Balmer decrement, since the Paschen lines experience less attenuation by dust.

Non-unity dust covering fractions cause a larger fraction of the observed Balmer emission to originate from less reddened OB associations, whereas the Paschen lines are less affected by this bias and include emission from dustier regions. Combined with the Paschen lines probing dusty regions that are optically thick to the Balmer lines, $E(B-V)\textrm{-PB}$ values are expected to be similar to or greater than their $E(B-V)\textrm{-BD}$ counterparts. This expectation is consistent with what is observed in Figure \ref{fig: Nebular Reddening; New Slit-Loss Corrections}, where two galaxies exhibit significantly higher $E(B-V)\textrm{-PB}$ values and no galaxies show significantly higher $E(B-V)\textrm{-BD}$ values. Additionally, composite Paschen reddenings are 45\% larger for the Pa$\alpha$-detected sample and 24\% larger for the Pa$\beta$-detected sample. Although the composite $E(B-V)\textrm{-BD}$ and $E(B-V)\textrm{-PB}$ values agree within one standard deviation, the elevated composite Paschen reddening values may reflect the combined effects of non-unity covering fractions and the Paschen lines tracing highly obscured star-forming regions, which together drive the composite $E(B-V)\textrm{-PB}$ values higher.

\section{Star-Formation Rates} \label{sec:SFRs}

Hydrogen recombination lines are crucial for constraining short-timescale SFRs. Balmer lines, however, may be biased toward less dusty star-forming regions, while Paschen lines probe all regions with recent star formation, including those that are optically thick in the Balmer lines. Consequently, the Paschen lines are expected to yield higher SFRs than Balmer lines. By comparing SFRs derived from dust-corrected H$\alpha$, Pa$\alpha$, and Pa$\beta$ luminosities, the extent of additional star formation revealed by the Paschen lines can be assessed. We can also compare these recombination line-based SFRs with UV-based SFRs, which trace star formation over longer timescales. These comparisons can yield insights into the recent star-formation histories of galaxies and the appropriate dust corrections to apply to the UV stellar continuum.

\subsection{Recombination Line SFRs}\label{subsec:Recombination SFR}

SFRs were calculated by dust-correcting the observed fluxes assuming the MW extinction curve, using $E(B-V)\textrm{-BD}$ for H$\alpha$ and $E(B-V)\textrm{-PB}$ for Pa$\alpha$ and Pa$\beta$. The dust-corrected luminosities were converted to SFRs using the following equations:

\begin{equation}
\mathrm{SFR(Pa\alpha)}[M_\odot\ \mathrm{yr^{-1}}]=C(\mathrm{Pa\alpha})\times L(\mathrm{Pa\alpha)[ergs\ s^{-1}}]
\label{eq:Paa SFR}
\end{equation}
\begin{equation}
\mathrm{SFR(Pa\beta)}[M_\odot\ \mathrm{yr^{-1}}]=C(\mathrm{Pa\beta})\times L(\mathrm{Pa\beta)[ergs\ s^{-1}}]
\label{eq:Pab SFR}
\end{equation}
\begin{equation}
\mathrm{SFR(H\alpha)}[M_\odot\ \mathrm{yr^{-1}}]=C(\mathrm{H\alpha})\times L(\mathrm{H\alpha)[ergs\ s^{-1}}].
\label{eq:Ha SFR}
\end{equation}

\noindent For these equations, $C(\rm Pa\alpha)=1.95\times10^{-41}$, $C(\rm Pa\beta)=3.84\times10^{-41}$, and $C(\rm H\alpha)=2.14\times 10^{-42}$, corresponding to a $Z_*=0.001$ (subsolar metallicity) BPASS model with a constant star-formation history and a \cite{Chabrier2003} IMF with an upper mass cutoff of $100M_\odot$ \citep{Reddy2022}.

\begin{figure}

\includegraphics[width=0.45\textwidth]{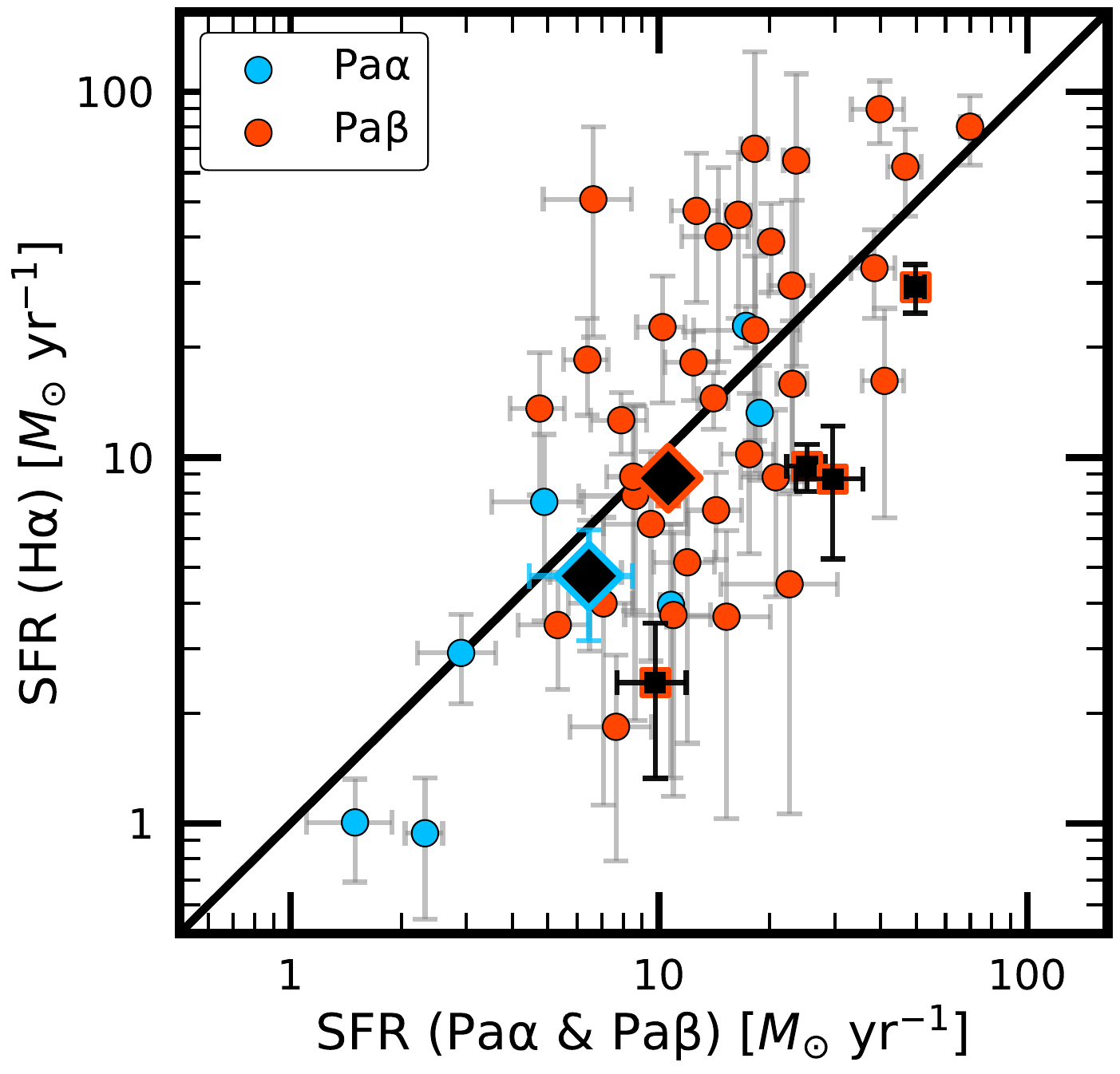}
\caption{H$\alpha$ SFRs plotted against Pa$\alpha$ and Pa$\beta$ SFRs. Blue and red points represent galaxies with Pa$\alpha$ and Pa$\beta$ detections, respectively, with diamonds indicating values measured from the composite spectra. Galaxies with one SFR value significantly greater than the other are marked with black squares. The solid line denotes equality between Paschen-line and H$\alpha$ SFRs. Four galaxies show significantly higher Paschen SFRs, while none have significantly larger H$\alpha$ SFRs.}
\label{fig: Recombination SFR Ratios}
\end{figure}

The Paschen-line and H$\alpha$ SFRs are shown in Figure \ref{fig: Recombination SFR Ratios}. Four galaxies exhibit Paschen SFRs significantly larger than their H$\alpha$ SFRs, including the two galaxies for which $E(B-V)\textrm{-PB}$ is significantly greater than $E(B-V)\textrm{-BD}$; no galaxies show significantly higher H$\alpha$ SFRs. The composite Pa$\alpha$ SFR is 36\% larger than the composite H$\alpha$ SFR, and the composite Pa$\beta$ SFR is 21\% larger than that derived from H$\alpha$. The Paschen- and Balmer-derived SFRs from the composite spectra agree within one standard deviation. This mirrors the results from the analysis of nebular reddening, where Balmer and Paschen estimates generally agree on average, but the Paschen lines can reveal significantly more star formation in specific cases. The same factors contributing to higher $E(B-V)\textrm{-PB}$ values, such as the Paschen lines tracing regions optically thick to the Balmer lines, also explain the larger Paschen SFRs. If $E(B-V)\textrm{-BD}$ is underestimated, the dust-corrected H$\alpha$ flux and corresponding SFR will be underestimated as well. Future studies that exclusively use Paschen lines for dust corrections and SFR measurements can more effectively trace star formation in the dustiest nebular regions compared to Pa$\alpha$/H$\alpha$ or Pa$\beta$/H$\alpha$.

\subsection{UV SFRs}\label{subsec:Paschen and UV SFR}

The Paschen lines provide the most robust measurements of recent star formation on timescales of $\approx\!10$\,Myr, enabling direct comparisons with SFRs inferred from the UV continuum, which traces star formation over longer timescales of $\sim\!100$\,Myr. The relationship between recombination line- and UV-based SFRs is dependent on a variety of factors, including the galaxy's star-formation history \citep{Erb2006, Guo2016, Emami2019, Faisst2019, Atek2022}, differences in attenuation toward H\,$\rm\text{\small II}$ regions versus toward older stellar populations \citep{Reddy2010, Reddy2015, Shivaei2015, Shivaei2018, Theios2019, Fetherolf2021}, the galaxy's IMF \citep{Leitherer1995, Boselli2009, Meurer2009, Pflamm2009}, the ionizing escape fraction \citep{Steidel2001, Shapley2006, Siana2007}, and binary stellar evolution \citep{Eldridge2012, Choi2017, Eldridge2017, Rezaee2023}. Here, we directly compare Paschen and UV-based SFRs, where the latter have been corrected for dust assuming the SMC extinction curve.

\begin{figure}

\includegraphics[width=0.45\textwidth]{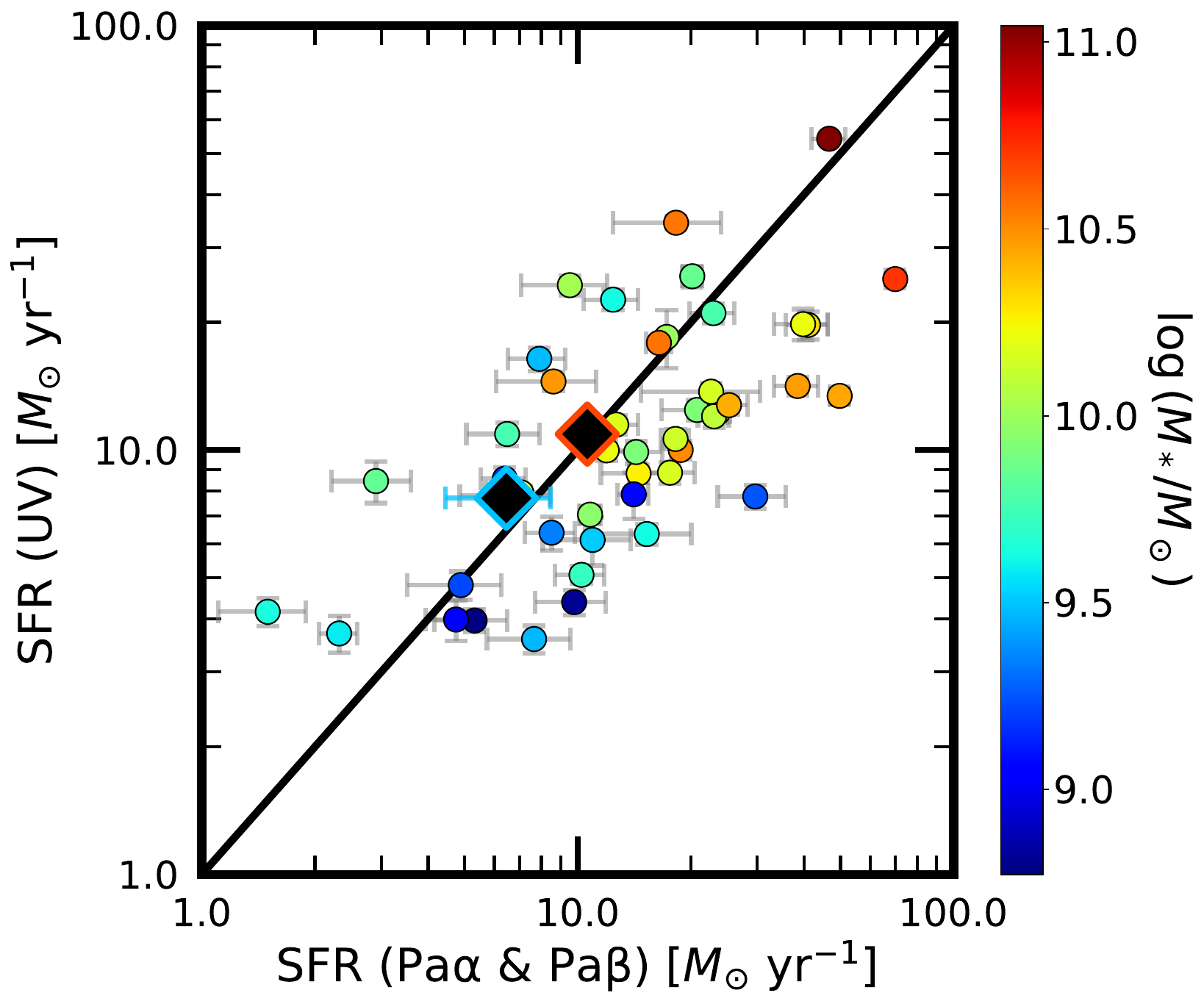}

\caption{Dust-corrected UV SFRs plotted against dust-corrected Paschen SFRs, color-coded by stellar mass. Composite Paschen SFRs and average UV SFRs for all 77 galaxies are shown as black diamonds. The solid black line denotes equality between the two SFR estimates.}
\label{fig: UV SFRs}

\end{figure}

To calculate UV SFRs, the flux density at 1600\,\r{A} was determined from the best-fit SED, then dust corrections were applied using the SED-determined reddening and assuming the SMC extinction curve. The UV SFR was calculated using the following equation:

\begin{equation}
\mathrm{SFR(UV)}[M_\odot\ \mathrm{yr^{-1}}]=C(\mathrm{UV})\times\nu L_\nu\mathrm{[ergs\ s^{-1}]},
\label{eq:UV SFR}
\end{equation}

\noindent where $C(\rm UV)=3.72\times10^{-44}$ for a $Z_*=0.001$ (subsolar metallicity) BPASS model with a constant star-formation history and an age of 100\,Myr. Figure \ref{fig: UV SFRs} shows UV SFRs plotted against Paschen SFRs dust-corrected using $E(B-V)\textrm{-PB}$. Of the 46 individual galaxies, 30 show Paschen SFRs greater than their UV SFRs, with Paschen values averaging 47\% higher for the individually-analyzed sample. Additionally, we do not observe increased scatter between Paschen and UV SFRs at lower stellar masses or SFRs, in contrast to previous studies \citep[e.g.,][]{Dominguez2015, Emami2019, Atek2022, Cleri2022, Reddy2023}. The larger Paschen SFRs and lack of scatter are likely due to the $3\sigma$-detection requirement for Pa$\alpha$ and Pa$\beta$. This requirement excludes galaxies with fainter Paschen lines, which are more likely to have Paschen SFRs that are more comparable to their UV SFRs. As a result, this S/N limit biases the individually-analyzed sample toward higher Paschen-to-UV SFR ratios. The limit also excludes lower-SFR and lower-mass galaxies, where greater scatter between the two SFR measurements is typically seen \citep[e.g.,][]{Dominguez2015, Emami2019, Atek2022, Cleri2022, Reddy2023}.

We also calculated average UV SFRs for all galaxies in the sample, including those without formal Paschen line detections. Comparing these UV SFRs to Paschen SFRs derived from the composite spectra, we find ratios of $\rm SFR(UV)/SFR(Pa\alpha)=1.19\pm0.35$ and $\rm SFR(UV)/SFR(Pa\beta)=1.03\pm0.12$. Both ratios are within one standard deviation of unity, indicating that the elevated Paschen SFRs observed in the individual galaxies are driven by the $3\sigma$-detection requirement. Future studies with deeper observations of low-SFR galaxies can further explore the consistency between Paschen and UV SFRs at the low end of the SFR range and assess whether the increased scatter reported in previous studies persists.

\section{Correlations with Galaxy Properties}\label{sec:Reddening Correlations}

In galaxies with non-unity dust covering fractions, the observed Balmer decrement will be biased toward less reddened OB associations. Additionally, some galaxies may contain regions so dust-enshrouded that they are opaque to the Balmer lines altogether. In such cases, Balmer-derived reddening measurements may become decoupled from the galaxy's total nebular reddening, while the Paschen-derived measurements remain more closely tied to the total reddening. Nebular reddening has been found to correlate with a number of other galaxy properties, including reddening of the stellar continuum \citep[e.g.,][]{Fanelli1988, Calzetti1994, Calzetti1997, Calzetti2000, Yoshikawa2010, Kreckel2013, Price2014, Reddy2015, Reddy2020, Shivaei2020, Fetherolf2021, Fetherolf2023}, stellar mass \citep[e.g.,][]{Gillespie2003, Brinchmann2004, Garn2010, Dominguez2013, Kashino2013, Zahid2017, Theios2019, Shivaei2020, Matharu2023, Lorenz2024, Maheson2024}, metallicity \citep[e.g.,][]{Asari2007, Garn2010, Xiao2012, Boselli2013, Zahid2014, Shivaei2020, Fetherolf2023, Lorenz2024, Maheson2024}, and SFR \citep[e.g.,][]{Hopkins2001, Afonso2003, Garn2010, Ly2012, Xiao2012, Dominguez2013, Zahid2013, Reddy2015, Shivaei2020, Lorenz2024}. By examining how the Balmer- and Paschen-derived reddening measurements vary with these galaxy properties, we can gain insight into the factors that influence the reddening values and their potential differences. Here, we investigate how $E(B-V)\textrm{-BD}$ and $E(B-V)\textrm{-PB}$ correlate with these properties.

\subsection{Reddening of the Stellar Continuum}\label{subsec:Stellar Reddening}

Comparisons between nebular reddening and reddening of the stellar continuum ($E(B-V)_\textrm{stars}$) offer direct insight into how dust affects the lines of sight toward young, massive stars embedded in their dusty birth clouds versus those toward the broader stellar population. The dustiest OB associations are faint in the UV continuum, due to the greater attenuation. As a result, the UV stellar continuum is dominated by less reddened OB associations, including those whose birth clouds have dissipated but whose surviving stars still contribute significantly to the continuum. For this reason, $E(B-V)_\textrm{stars}$ is expected to be lower than nebular reddening. Comparing $E(B-V)\textrm{-BD}$ and $E(B-V)\textrm{-PB}$ to $E(B-V)_\textrm{stars}$ can provide insight into how dust is distributed between the youngest and slightly older stars.

The relationship between reddening of nebular lines and of the stellar continuum has been the focus of many previous studies. The exact relation between $E(B-V)_\textrm{stars}$ and nebular reddening depends on the choice of assumed stellar attenuation curve. \cite{Calzetti2000} reported an $E(B-V)_\textrm{stars}/E(B-V)\textrm{-BD}$ ratio of 0.44, which has been found to hold in general in high-redshift star-forming galaxies when assuming the Calzetti extinction curve to dust-correct the stellar continuum \citep[e.g.,][]{Schreiber2009, Yoshikawa2010, Mancini2011, Wuyts2011, Price2014, Tran2015, Buat2018, Reddy2020, Shivaei2020}. In a sample of $z\sim2$ star-forming galaxies, \cite{Reddy2020} found that assuming the SMC extinction curve lowers the ratio to 0.23, because the steeper slope of the SMC curve yields smaller $E(B-V)_\textrm{stars}$ values \citep{Theios2019, Reddy2020}. Other studies have found that this ratio tends to decrease with SFR, consistent with a scenario in which a moderately reddened stellar population dominates the continuum, while a second, dustier stellar population begins to dominate both the bolometric and nebular line luminosities at higher SFRs \citep[e.g.,][]{Kreckel2013, Price2014, Reddy2015, DeBarros2016, Shivaei2020, Fetherolf2021, Lorenz2024}.

\begin{figure*}

\includegraphics[width=0.95\textwidth]{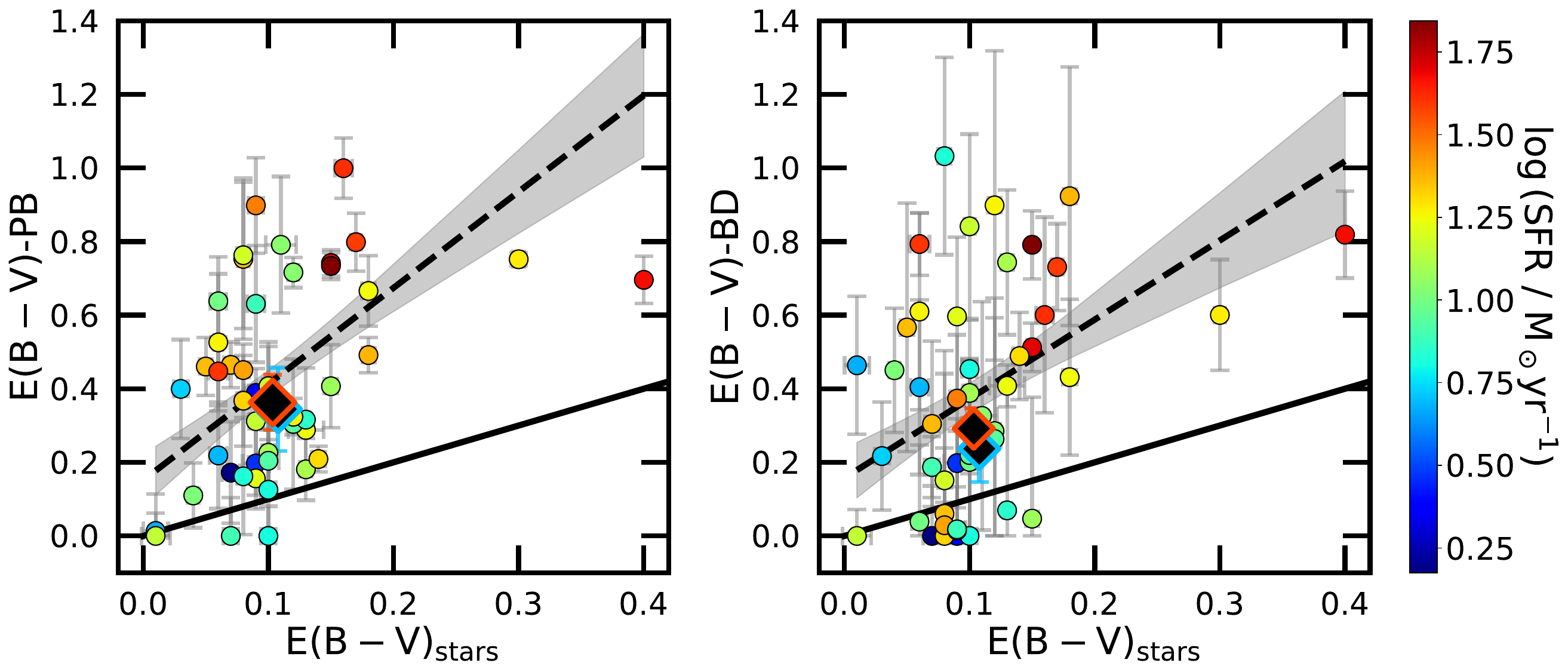}
\caption{Nebular reddening from the Paschen-based decrements (left) and Balmer decrement (right) plotted against SED-derived reddening of the stellar continuum, color-coded by dust-corrected Paschen SFR. Composite nebular reddenings and average $E(B-V)_\textrm{stars}$ values for all 77 galaxies are shown as black diamonds. The solid black lines indicate equality between reddenings, while the dashed lines and shaded areas represent the best-fit lines and their 68\% confidence intervals. The two outlier points with the largest $E(B-V)_\textrm{stars}$ values correspond to the galaxies with the highest stellar masses in the Pa$\alpha$- and Pa$\beta$-detected subsamples.}
\label{fig: Stellar and Nebular Reddening}
\end{figure*}

We compared Balmer- and Paschen-derived nebular reddening with the reddening derived from the best-fit SED. Figure \ref{fig: Stellar and Nebular Reddening} shows $E(B-V)\textrm{-BD}$ and $E(B-V)\textrm{-PB}$ plotted against $E(B-V)_\textrm{stars}$, color-coded by dust-corrected Paschen SFR. Of the 46 galaxies, 34 show $E(B-V)\textrm{-BD}>E(B-V)_\textrm{stars}$, while 43 show $E(B-V)\textrm{-PB}>E(B-V)_\textrm{stars}$. Spearman tests show positive correlations between $E(B-V)_\textrm{stars}$ and both $E(B-V)\textrm{-PB}$ (p-value = 0.013; $\rho=0.37$) and $E(B-V)\textrm{-BD}$ (p-value = 0.016; $\rho=0.35$). The slopes of the best-fit lines are significantly positive as well, with confidence levels of $7.00\sigma$ for $E(B-V)\textrm{-PB}$ and $2.67\sigma$ for $E(B-V)\textrm{-BD}$. The composite spectra yield an average $E(B-V)_\textrm{stars}/E(B-V)\textrm{-PB}$ ratio of $0.31\pm0.09$ for the Pa$\alpha$-detected sample and $0.29\pm0.06$ for the Pa$\beta$-detected sample. For $E(B-V)_\textrm{stars}/E(B-V)\textrm{-BD}$, the ratios are $0.46\pm0.15$ and $0.35\pm0.07$ for the Pa$\alpha$- and Pa$\beta$-detected samples, respectively. The color-coding in Figure \ref{fig: Stellar and Nebular Reddening} indicates that galaxies with $E(B-V)\textrm{-PB}$ or $E(B-V)\textrm{-BD}$ values that substantially exceed $E(B-V)_\textrm{stars}$ tend to have higher SFRs.

$E(B-V)\textrm{-BD}$ and $E(B-V)\textrm{-PB}$ both show positive correlations with $E(B-V)_\textrm{stars}$, indicating that OB associations are contributing significantly to the UV continuum for all galaxies in the sample. The presence of this trend in both Balmer- and Paschen-derived reddening shows that the Balmer and Paschen lines each trace this relation with stellar continuum emission. We also find that $E(B-V)\textrm{-BD}$ and $E(B-V)\textrm{-PB}$ exceed $E(B-V)_\textrm{stars}$ in most galaxies in the sample. This is consistent with previous studies, which attribute this offset to nebular emission including OB associations along dustier lines of sight than those contributing to the far-UV continuum. The $E(B-V)_\textrm{stars}/E(B-V)\textrm{-PB}$ ratios from the composite spectra are consistent within $1\sigma$ with the value of 0.23 reported when assuming the SMC extinction curve, and the $E(B-V)_\textrm{stars}/E(B-V)\textrm{-BD}$ composite ratios are consistent within $2\sigma$. The lower $E(B-V)_\textrm{stars}/E(B-V)\textrm{-PB}$ ratios arise due to the higher Paschen-derived nebular reddenings measured from the composite spectra relative to those from the Balmer decrement. It is important to note that $E(B-V)_\textrm{stars}$ depends on the extinction curve assumed in the SED modeling, and was derived here adopting a fixed SMC curve. Assuming a shallower curve, such as the Calzetti curve, would yield larger $E(B-V)_\textrm{stars}$ values on average, increasing the stellar-to-nebular reddening ratios and potentially affecting the strength of the correlation between nebular reddening and reddening of the stellar continuum.

\subsection{Stellar Mass}\label{subsec:Stellar Mass}

We next examined correlations between the nebular reddenings and SED-derived stellar mass. More massive galaxies with extended star-formation histories have produced more dust than less massive systems. Additionally, the deeper gravitational potential wells of higher-mass galaxies retain dust more efficiently. For these reasons, higher-mass galaxies are expected to exhibit greater dust attenuation, as observed in previous studies \citep[e.g.,][]{Giovanelli1995, Tully1998, Gillespie2003, Kauffmann2003, Brinchmann2004, Calura2008, DaCunha2010, Garn2010, Dominguez2013, Kashino2013, Zahid2017, Theios2019, Matharu2023, Maheson2024}. Here, we investigate how well $E(B-V)\textrm{-BD}$ and $E(B-V)\textrm{-PB}$ trace this association between stellar mass and attenuation.

\begin{figure*}

\includegraphics[width=0.95\textwidth]{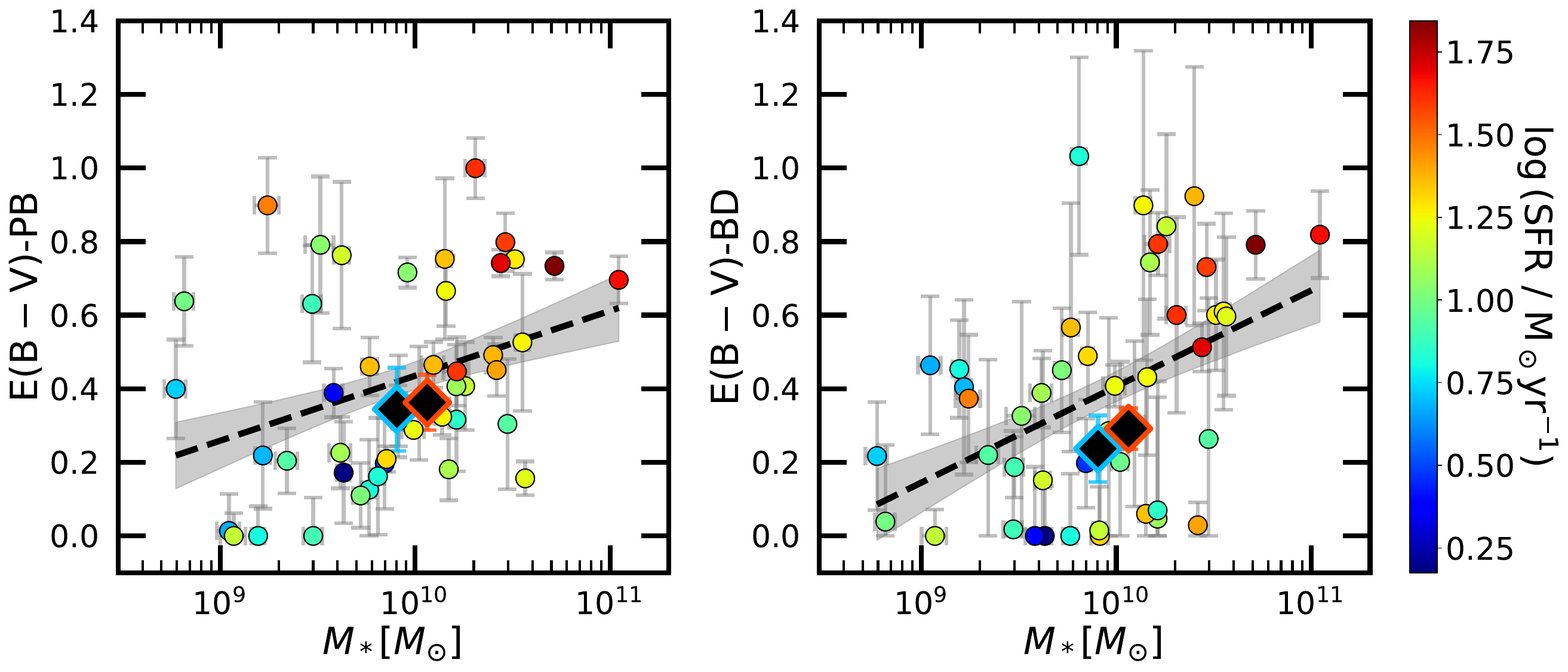}
\caption{Paschen-derived (left) and Balmer decrement-derived (right) nebular reddening plotted against SED-derived stellar mass, color-coded by dust-corrected Paschen SFR. Composite nebular reddening values and average stellar masses for all 77 galaxies are shown as black diamonds. The dashed lines indicate the best-fit lines, and the shaded areas represent the 68\% confidence intervals. Both $E(B-V)\textrm{-PB}$ and $E(B-V)\textrm{-BD}$ exhibit significant positive correlations with stellar mass.}
\label{fig: Reddening and Stellar Mass}
\end{figure*}

The measured nebular reddening values and stellar masses are shown in Figure \ref{fig: Reddening and Stellar Mass}. Spearman tests reveal positive correlations for both measures of nebular reddening, with p-values of 0.010 ($\rho=0.37$) for $E(B-V)\textrm{-PB}$ and $7.68\times10^{-4}$ ($\rho=0.48$) for $E(B-V)\textrm{-BD}$. The slopes of the best-fit lines are also significantly positive, with confidence levels of $5.59\sigma$ for $E(B-V)\textrm{-PB}$ and $4.62\sigma$ for $E(B-V)\textrm{-BD}$.

Stellar mass shows a clear correlation with both $E(B-V)\textrm{-BD}$ and $E(B-V)\textrm{-PB}$, indicating that both reddening measures effectively trace the higher dust content associated with more massive galaxies. Because dust optical depths increase with stellar mass, one might expect more massive galaxies to contain lines of sight so heavily obscured that they are opaque to the Balmer lines, causing $E(B-V)\textrm{-BD}$ to diverge from $E(B-V)\textrm{-PB}$ and potentially weakening the correlation between $E(B-V)\textrm{-BD}$ and stellar mass. However, we still observe a strong trend between $E(B-V)\textrm{-BD}$ and stellar mass, because the optical depths along all the lines of sight toward OB associations tend to increase with stellar mass, not only those that are optically thick in the Balmer lines. Consequently, even if $E(B-V)\textrm{-BD}$ diverges from $E(B-V)\textrm{-PB}$ and underestimates the true average attenuation in massive galaxies, its correlation with stellar mass remains largely preserved.

\subsection{O3N2 Metallicity}\label{subsec:Metallicity}

We next examined the relationships between the nebular reddenings and O3N2 metallicity \citep{Pettini2004}. For this analysis, 44 galaxies with coverage of the $\rm[O\,\text{\small III}]\lambda5008$ and $\rm[N\,\text{\small II}]\lambda6585$ lines in the MOSDEF observations were selected. Unlike stellar mass and $E(B-V)_\textrm{stars}$, O3N2 metallicity is derived from nebular emission lines that originate from H\,$\rm\text{\small II}$ regions. A positive correlation between metallicity and reddening is expected, as a higher metal content in the ISM facilitates the formation of dust grains \citep[e.g.,][]{Issa1990, Heckman1998, Leitherer1999, Calzetti2001, Madden2006, Asari2007, Munoz2009, Garn2010, Reddy2010, Xiao2012, Boselli2013, Zahid2014, Maheson2024}.

\begin{figure*}

\includegraphics[width=0.95\textwidth]{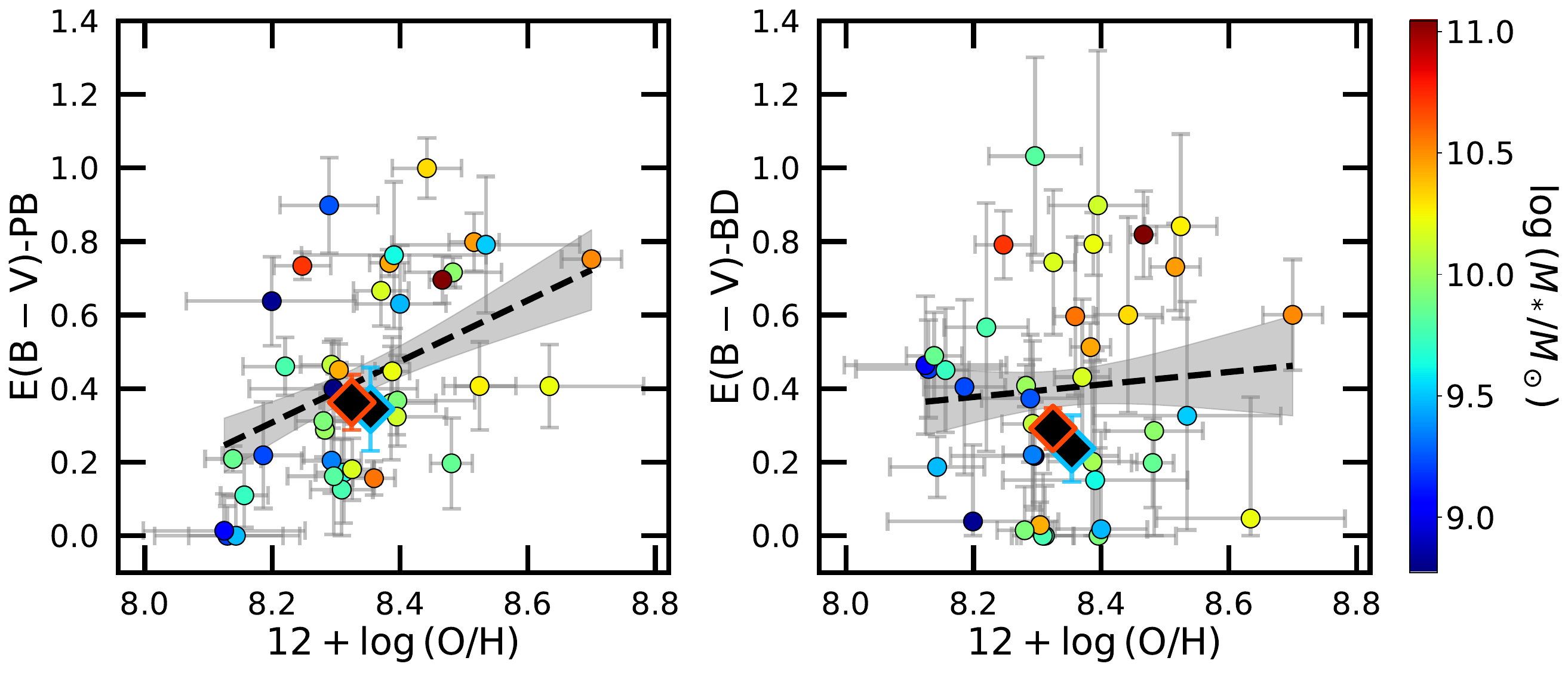}

\caption{Paschen-derived (left) and Balmer decrement-derived (right) nebular reddening plotted against O3N2 metallicity for 44 galaxies with coverage of $\rm[O\,\text{\small III}]\lambda5008$ and $\rm[N\,\text{\small II}]\lambda6585$, color-coded by stellar mass. Composite nebular reddening values and average metallicities for all galaxies with coverage of the required emission lines are shown as black diamonds. The dashed lines indicate the best-fit lines, and the shaded areas represent the 68\% confidence intervals. A significant positive correlation is observed between metallicity and $E(B-V)\textrm{-PB}$, while no correlation is found for $E(B-V)\textrm{-BD}$.}
\label{fig: Nebular Reddening and Metallicity}

\end{figure*}

Figure \ref{fig: Nebular Reddening and Metallicity} shows the Paschen- and Balmer-derived nebular reddenings plotted against O3N2 metallicity. We find that $E(B-V)\textrm{-PB}$ correlates with metallicity, while $E(B-V)\textrm{-BD}$ does not show a significant trend. Spearman tests yield p-values of $8.72\times10^{-4}$ ($\rho=0.52$) for $E(B-V)\textrm{-PB}$ and 0.597 ($\rho=0.09$) for $E(B-V)\textrm{-BD}$. The slope of the best-fit line is positive with a confidence level of $4.37\sigma$ for $E(B-V)\textrm{-PB}$, but only $0.61\sigma$ for $E(B-V)\textrm{-BD}$.

The expected correlation is observed with $E(B-V)\textrm{-PB}$, but not with $E(B-V)\textrm{-BD}$. It is possible that non-unity dust covering fractions can explain the trends observed here, as they can bias $E(B-V)\textrm{-BD}$ toward values from relatively unreddened OB associations. In such cases, the increased dust content associated with higher metallicity decreases the fraction of Balmer emission observed from more heavily reddened regions. As a result, $E(B-V)\textrm{-BD}$ values decrease at high metallicity, weakening the observed correlation. In contrast, Paschen-based decrements remain more sensitive to heavily reddened OB associations and thus retain a stronger correlation with metallicity. Because the dust content along all lines of sight toward OB associations is expected to increase with metallicity, some correlation between $E(B-V)\textrm{-BD}$ and metallicity would still be expected. However, the observed O3N2 index is also subject to bias from non-unity dust covering fractions, which further weakens the correlation between $E(B-V)\textrm{-BD}$ and metallicity.

Because the O3N2 index uses nebular emission lines at wavelengths similar to those of H$\alpha$ and H$\beta$, one would expect metallicity and $E(B-V)\textrm{-BD}$ to be similarly biased toward the same less reddened nebular regions. To investigate this, a non-unity dust covering fraction model similar to that used in Section \ref{subsec:Covering Fractions} was applied, to evaluate its effects on measured O3N2 values. We assumed that the attenuated emission comes from a high-metallicity region, while the unattenuated emission originates from a low-metallicity region. To model the flux, in each region the emitted $\rm[O\,\text{\small III}]/H\beta$ and $\rm[N\,\text{\small II}]/H\alpha$ ratios were fixed to values observed in the sample. These values were taken from the highest- ($12+\log{(\mathrm{O/H})}=8.52$) and lowest-metallicity ($12+\log{(\mathrm{O/H})}=8.14$) galaxies with ${\geq}3\sigma$ detections of all four lines. For the high-metallicity region, $\log{(\rm[O\,\text{\small III}]/H\beta)}=0.11$ and $\log{(\rm[N\,\text{\small II}]/H\alpha)}=-0.54$; for the low-metallicity region, $\log{(\rm[O\,\text{\small III}]/H\beta)}=0.68$ and $\log{(\rm[N\,\text{\small II}]/H\alpha)}=-1.18$. O3N2 metallicity was then calculated across a range of covering fractions and attenuation values using the combined emission of both regions.

\begin{figure*}

\includegraphics[width=0.95\textwidth]{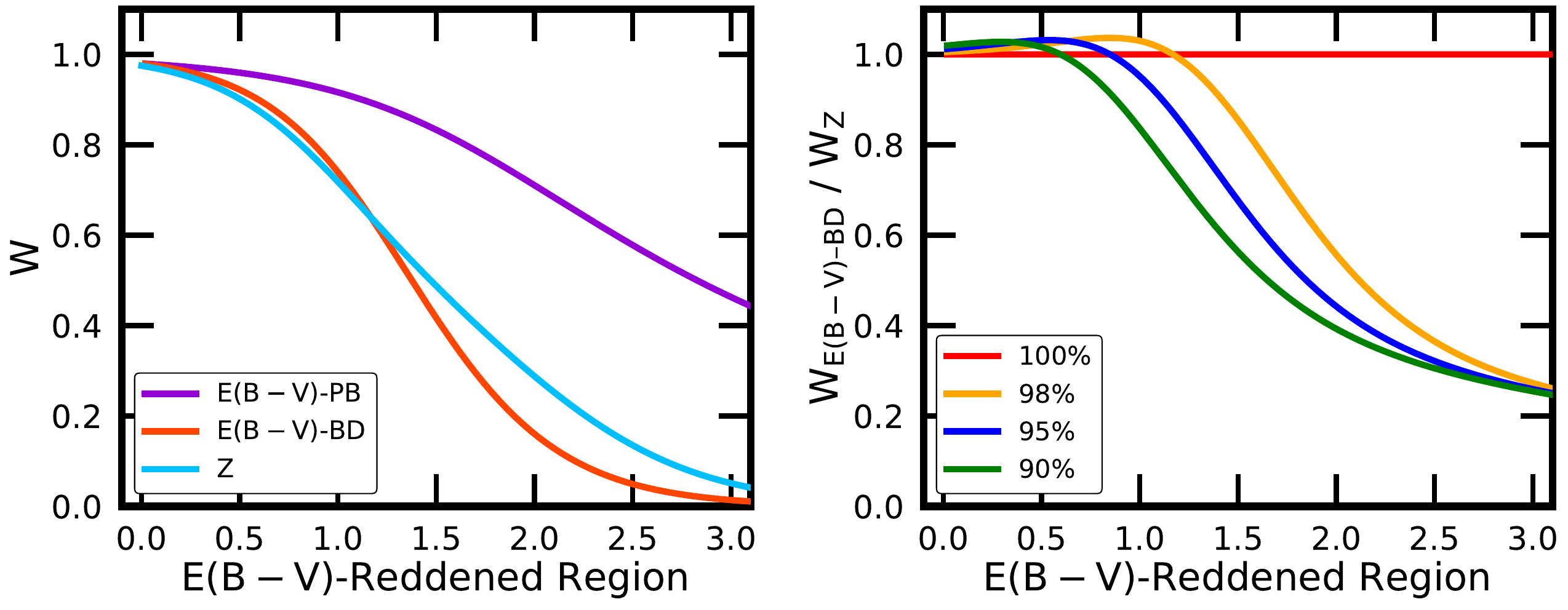}
\caption{The left panel shows the weighting of measured $E(B-V)\textrm{-PB}$, $E(B-V)\textrm{-BD}$, and O3N2 metallicity values toward the high-dust, high-metallicity region for a 98\% covering fraction, plotted against the reddening of the attenuated component. The right panel shows the ratio of the $E(B-V)\textrm{-BD}$ weighting to O3N2 metallicity weighting for covering fractions of 100\%, 98\%, 95\%, and 90\%, also plotted against the reddening of the attenuated component. At lower attenuation, the measured $E(B-V)\textrm{-BD}$ and metallicity values are similarly weighted between the two regions, while at higher attenuation $E(B-V)\textrm{-BD}$ becomes increasingly more weighted toward the unattenuated, low-metallicity region relative to O3N2 metallicity. $E(B-V)\textrm{-PB}$ remains consistently more weighted toward the attenuated, high-metallicity region.}
\label{fig: Non-unity Covering Weights}
\end{figure*}

To quantify the relative contribution of each region to the measured metallicity, we defined the following weighting parameter:

\begin{equation}
W_Z=\frac{Z_{\rm obs}-Z_{\rm Unattenuated}}{Z_{\rm Attenuated}-Z_{\rm Unattenuated}},
\label{eq:Weighting Equation}
\end{equation}

\noindent where $Z_{\rm obs}$ is the metallicity measured from the blended emission of both regions, while $Z_{\rm Attenuated}$ and $Z_{\rm Unattenuated}$ are the metallicities of the attenuated and unattenuated regions, respectively. Here, $W_Z=1$ indicates that the measured O3N2 fully reflects the attenuated region (with a higher metallicity) and $W_Z=0$ indicates that the measured O3N2 is completely biased to the unattenuated region (with a lower metallicity). Using this formalism, we also defined weighting parameters for the measured $E(B-V)\textrm{-BD}$ ($W_{E(B-V)\textrm{-BD}}$) and $E(B-V)\textrm{-PB}$ ($W_{E(B-V)\textrm{-PB}}$) values in this non-unity covering fraction model:

\begin{equation}
W_{E(B-V)}=\frac{E(B-V)_{\rm obs}}{E(B-V)_{\rm Attenuated}},
\label{eq:Reddening Weighting Equation}
\end{equation}

\noindent where $E(B-V)_{\rm obs}$ is the observed reddening and $E(B-V)_{\rm Attenuated}$ is the reddening of the attenuated region. As with $W_Z$, $W_{E(B-V)}=1$ indicates that the measured reddening matches that of the attenuated region, while $W_{E(B-V)}=0$ indicates that the observed emission is entirely from the unattenuated region and thus yields zero reddening. The $W_Z$, $W_{E(B-V)\textrm{-BD}}$, and $W_{E(B-V)\textrm{-PB}}$ values for a 98\% covering fraction are shown in the left panel of Figure \ref{fig: Non-unity Covering Weights}, and the $W_Z/W_{E(B-V)\textrm{-BD}}$ ratios for covering fractions of 100\%, 98\%, 95\%, and 90\% are shown in the right panel.

In the right panel of Figure \ref{fig: Non-unity Covering Weights}, $E(B-V)\textrm{-BD}$ and O3N2 metallicity show similar weighting between the two regions at low attenuation for all covering fractions. However, as attenuation increases under non-unity covering fractions, $E(B-V)\textrm{-BD}$ becomes increasingly more weighted toward the low-dust, low-metallicity region compared to O3N2 metallicity. As a result, although both diagnostics are biased toward the unreddened region at high attenuation, the effect is more significant for the Balmer decrement. This differential bias weakens the correlation between $E(B-V)\textrm{-BD}$ and metallicity. In contrast, Paschen-derived reddening remains less biased toward the low-dust region across all attenuation levels. Thus, even if O3N2 metallicity is underestimated due to non-unity dust covering fractions, its correlation with $E(B-V)\textrm{-PB}$ remains robust.

Understanding why the measured O3N2 metallicity is less biased toward the dust-free region than $E(B-V)\textrm{-BD}$ under non-unity covering fractions is crucial, especially given that both rely on nebular lines that are close in wavelength and should therefore experience similar attenuation. $E(B-V)\textrm{-BD}$ traces the deviation of the observed Balmer decrement from its intrinsic dust-free value, since attenuation increases the observed ratio. As shown in Section \ref{subsec:Covering Fractions}, under non-unity covering fractions the observed Balmer decrement begins to decrease at higher attenuation levels, even though the decrement from the attenuated region continues to increase. This discrepancy causes the measured $E(B-V)\textrm{-BD}$ to diverge more significantly from the true reddening of the dusty component at higher attenuation. In contrast, the intrinsic O3N2 value of the attenuated region is determined solely by its metallicity and is therefore fixed and independent of attenuation. As a result, although increasing attenuation biases the measured O3N2 value toward that of the unreddened region, the discrepancy between measured O3N2 and intrinsic O3N2 of the reddened region increases more gradually with attenuation than the corresponding discrepancy for the Balmer decrement. In other words, the measured O3N2 and Balmer decrement both become biased toward the dust-free region at high attenuation, but the bias is more significant for the Balmer decrement because its value in the reddened region increases with attenuation while the O3N2 value is independent of attenuation. Consequently, at high attenuation, the observed $E(B-V)\textrm{-BD}$ is more strongly weighted toward the unattenuated region than O3N2 metallicity. Because $E(B-V)\textrm{-PB}$ is less biased toward the unattenuated region than $E(B-V)\textrm{-BD}$ across all attenuation levels, $E(B-V)\textrm{-PB}$ more accurately captures the correlation between nebular reddening and metallicity.

\subsection{UV SFR}\label{subsec:UV SFR}

Next, we investigated correlations between the nebular reddenings and UV SFR. Previous works have found that higher rates of star formation are associated with greater levels of dust attenuation \citep[e.g.,][]{Wang1996, Adelberger2000, Hopkins2001, Afonso2003, Reddy2006, Garn2010, Reddy2010, Ly2012, Sobral2012, Xiao2012, Dominguez2013, Zahid2013, Reddy2015}, so a correlation between UV SFR and reddening is expected. UV SFR is particularly sensitive to dust corrections applied to the stellar continuum, and examining its relationship with nebular reddening can reveal how dust-corrected UV luminosities are affected by non-unity dust covering fractions.

\begin{figure*}

\includegraphics[width=0.95\textwidth]{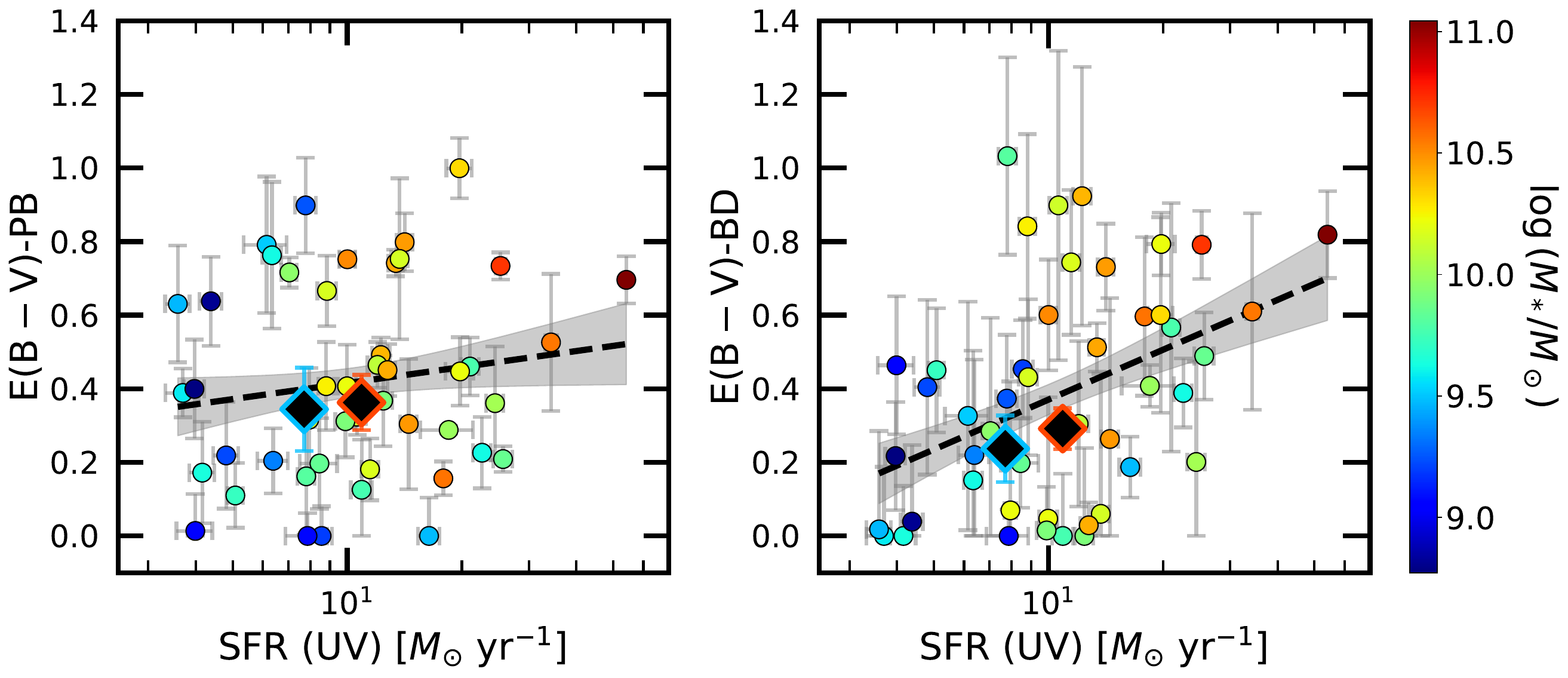}
\caption{Paschen-derived (left) and Balmer decrement-derived (right) nebular reddening plotted against UV SFR, color-coded by stellar mass. Composite nebular reddening values and average UV SFRs for all 77 galaxies are shown as black diamonds. The dashed lines indicate the best fit lines, and the shaded areas represent the 68\% confidence intervals. A significant positive correlation is found for $E(B-V)\textrm{-BD}$, but not for $E(B-V)\textrm{-PB}$.}
\label{fig: Reddening and UV SFRs}
\end{figure*}

Figure \ref{fig: Reddening and UV SFRs} shows the nebular reddening values plotted against UV SFRs. We find a significant positive correlation between $E(B-V)\textrm{-BD}$ and UV SFR, with a Spearman p-value of $4.86\times10^{-3}$ ($\rho=0.41$). The slope of the best-fit line is significantly positive with a confidence level of $4.03\sigma$. In contrast, $E(B-V)\textrm{-PB}$ shows no significant correlation with UV SFR, with a Spearman test returning a p-value of 0.368 ($\rho=0.14$). The slope of the best-fit line is positive with a confidence level of $2.21\sigma$.

We observe that UV SFR correlates with $E(B-V)\textrm{-BD}$, but not with $E(B-V)\textrm{-PB}$. Because SFR is generally correlated with dust attenuation, it is initially surprising that $E(B-V)\textrm{-PB}$ does not show such a trend. However, this behavior can be explained by non-unity dust covering fractions biasing the UV SFRs to less dusty regions, much like how the Balmer decrement is biased to less dusty regions relative to ratios including the Paschen lines. The best-fit SEDs assume a unity covering fraction. When the covering fraction is less than unity, the observed continuum emission, particularly in the UV, will preferentially come from less attenuated regions and the UV emission will reflect a smaller fraction of the total SFR as dustiness increases. In the dustiest galaxies, where many OB associations are heavily obscured by dust, the SED fit will underestimate the intrinsic UV luminosity and therefore underestimate SFR. In these cases, the Paschen lines still return the high nebular reddening values associated with these obscured regions, but the SED yields lower SFR estimates, weakening the correlation between reddening and UV SFR. In contrast, the Balmer lines are similarly biased toward less attenuated regions and thus yield lower reddening values. Under non-unity dust covering fractions, the relatively unreddened OB associations that dominate the observed Balmer emission also contribute significantly to the UV continuum. Thus, a correlation between UV SFR and Balmer reddening is still observed.

Although we find that non-unity dust covering fractions can bias estimated intrinsic UV luminosities toward lower values, in Section \ref{subsec:Paschen and UV SFR} we found that Paschen and UV SFRs agree on average across the full sample. This apparent discrepancy arises because non-unity dust covering fractions most strongly bias the UV luminosities of the dustiest galaxies. In these dustiest systems, UV luminosities are significantly underestimated, weakening the correlation between UV SFR and $E(B-V)\textrm{-PB}$ and contributing to the higher Paschen SFRs relative to UV SFRs observed in the individually-analyzed sample. In contrast, galaxies outside the individually-analyzed sample generally exhibit lower attenuation levels, and thus their UV luminosities are relatively less biased. As a result, when averaged over the full sample, the Paschen and UV SFRs yield similar values.

\subsection{Paschen SFR}\label{subsec:Paschen SFR}

We next investigated the correlation between Balmer decrement-derived nebular reddening and SFR determined from dust-corrected Paschen luminosity. The Paschen lines are robust tracers of dust-obscured star formation, particularly compared to the Balmer lines. However, many previous studies relied exclusively on Balmer lines to constrain star formation. Additionally, future surveys may have access only to Balmer lines and not Paschen. For this reason, it is important to assess whether the Balmer decrement is sensitive to star formation as measured from the Paschen lines.

\begin{figure}

\includegraphics[width=0.45\textwidth]{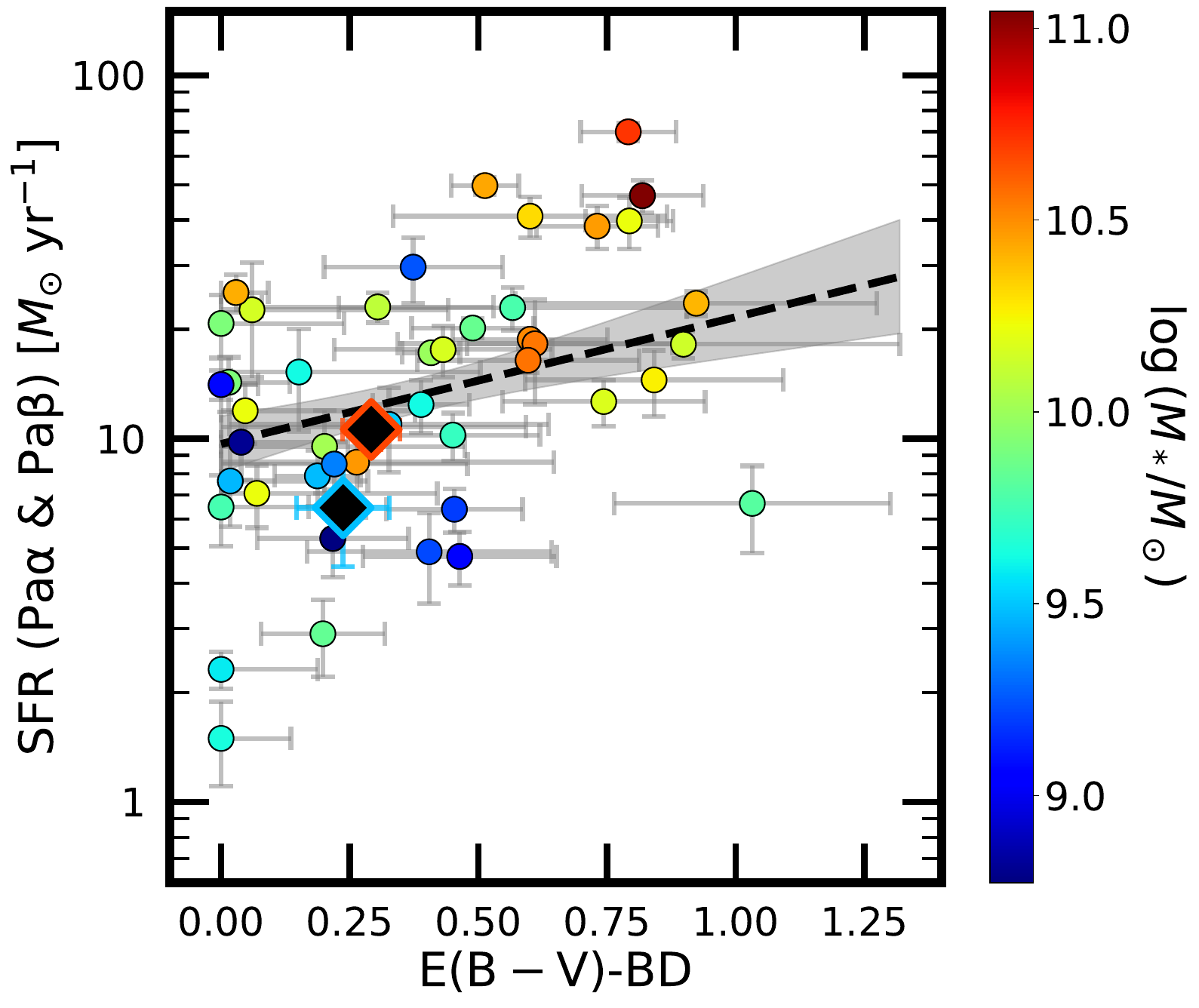}
\caption{Paschen SFR plotted against Balmer decrement-derived nebular reddening, color-coded by stellar mass. Nebular reddening and Paschen SFR values derived from the composite spectra are shown as black diamonds. The dashed line indicates the best fit line, and the shaded area represents the 68\% confidence interval. A significant positive correlation is found, indicating that the Balmer decrement traces star formation as probed by the Paschen lines.}
\label{fig: Balmer Reddening and Paschen SFR}
\end{figure}

Figure \ref{fig: Balmer Reddening and Paschen SFR} shows the Paschen SFRs plotted against Balmer reddening values. We find a significant positive correlation between $E(B-V)\textrm{-BD}$ and Paschen SFR, with a Spearman p-value of $1.39\times10^{-3}$ ($\rho=0.46$). The slope of the best-fit line is also significantly positive with a confidence level of $4.14\sigma$.

We observe that Paschen SFRs strongly correlate with $E(B-V)\textrm{-BD}$. The following equation shows the best-fit relation between Paschen SFR and $E(B-V)\textrm{-BD}$:

\begin{equation}
\log{(\mathrm{SFR}/M_\odot\ \mathrm{yr^{-1}})}=0.348\times E(B-V)\textrm{-BD}+0.986\ .
\label{eq:Paschen SFR and Balmer Reddening}
\end{equation}

\noindent This correlation indicates that, despite missing emission from the dustiest nebular regions, the Balmer lines still reliably trace obscured star formation. This result is consistent with the findings of Section \ref{subsec:Recombination SFR}, where we found that dust-corrected Balmer and Paschen SFRs agree on average. The loss of Balmer emission from heavily obscured regions is insufficient to weaken the correlation between reddening and Paschen SFR.

\subsection{Predictions of Nebular Reddening Differences}\label{subsec:Reddening Differences}

\begin{figure*}

\includegraphics[width=0.95\textwidth]{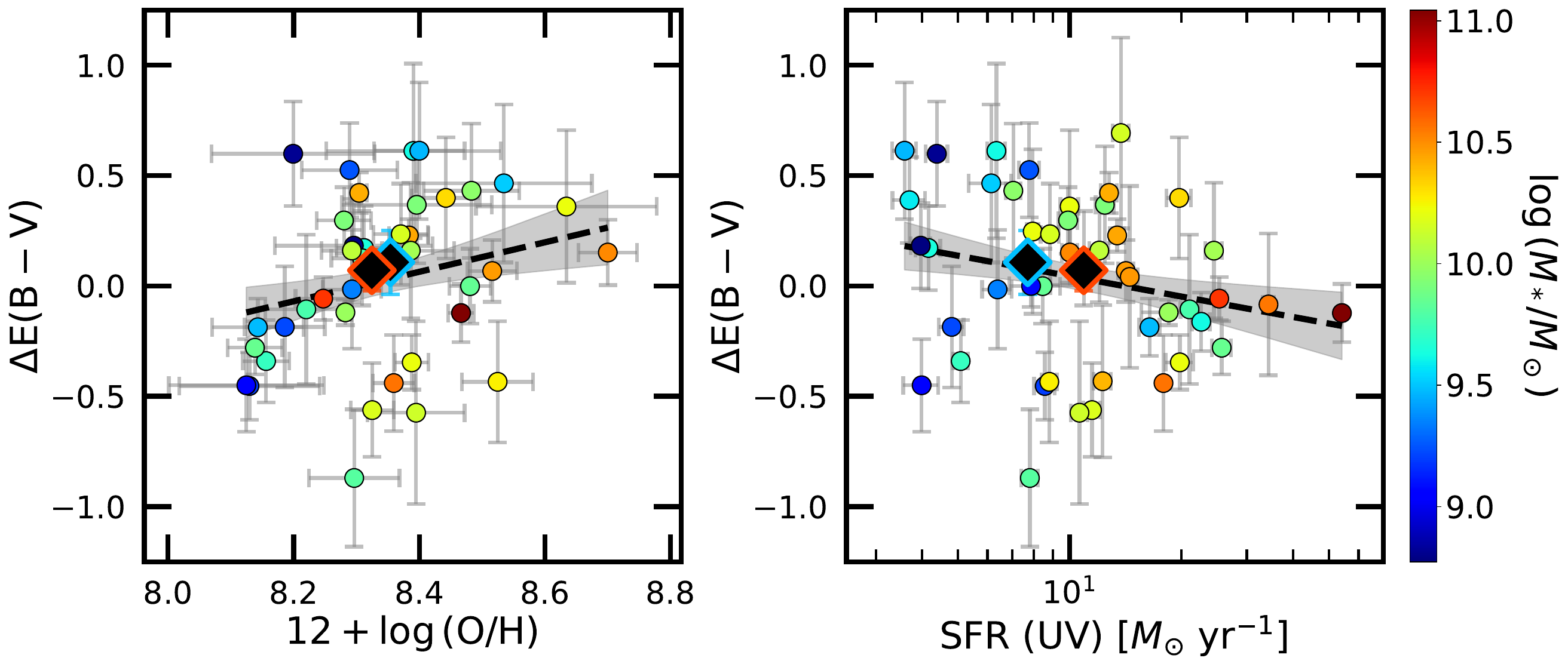}
\caption{$\Delta E(B-V)$ plotted against O3N2 metallicity (left) and UV SFR (right), color-coded by stellar mass. Composite nebular reddening values and average metallicities/UV SFRs for the full sample are shown as black diamonds. The dashed lines indicate the best-fit lines, and the shaded areas represent the 68\% confidence intervals. The correlations suggest that $E(B-V)\textrm{-PB}$ tends to exceed $E(B-V)\textrm{-BD}$ in galaxies with higher metallicities and lower UV SFRs.}
\label{fig: Reddening Differences}
\end{figure*}

Although the Balmer and Paschen lines yield nebular reddening and SFR values that agree on average, it is useful to formulate a method for estimating Paschen-based values when only the Balmer lines are available. Since only $E(B-V)\textrm{-PB}$ correlates with O3N2 metallicity and only $E(B-V)\textrm{-BD}$ correlates with UV SFR, these two properties can be used to estimate the difference between the two reddening measures. Figure \ref{fig: Reddening Differences} shows the difference between the two reddenings ($\Delta E(B-V)=E(B-V)\textrm{-PB}-E(B-V)\textrm{-BD}$) plotted against O3N2 metallicity and UV SFR. Spearman tests return p-values of 0.050 ($\rho=0.32$) for metallicity and 0.065 ($\rho=-0.27$) for UV SFR, indicating weak correlations such that $\Delta E(B-V)$ tends to increase with higher metallicity and lower UV SFR. The following equations present the best-fit relations for estimating $\Delta E(B-V)$ using O3N2 metallicity and UV SFR:

\begin{equation}
\Delta E(B-V)=0.672\times[12+\log{(\mathrm{O/H})}]-5.582
\label{eq:Reddening Difference and Metallicity}
\end{equation}
\begin{equation}
\Delta E(B-V)=-0.308\times\log{(\mathrm{SFR}/M_\odot\ \mathrm{yr^{-1}})}+0.353
\label{eq:Reddening Difference and SFR}
\end{equation}
\begin{multline}
\Delta E(B-V)=0.801\times[12+\log{(\mathrm{O/H})}]\\-0.308\times\log{(\mathrm{SFR}/M_\odot\ \mathrm{yr^{-1}})}-6.311\ .
\end{multline}

\noindent However, we caution that these relations are based on \textit{marginal} correlations between $\Delta E(B-V)$, O3N2 metallicity, and UV SFR. Future studies can refine these relations by using multiple Paschen lines to measure reddening in each galaxy, allowing for more effective tracing of star formation in the dustiest regions and enabling an investigation of how the correlations with O3N2 metallicity and UV SFR evolve with these new reddenings.

Table \ref{tab:appendix_table} in Appendix \ref{sec:Appendix Correlations} summarizes the results of the statistical tests performed in Section \ref{sec:Reddening Correlations}.

\section{Conclusions} \label{sec:Conclusion}

We present our analysis of 77 galaxies at $z=1.43-2.65$ observed in the MOSDEF and FRESCO surveys, using Balmer and Paschen recombination lines to compare nebular reddening and SFRs. Nebular and continuum emission maps from the FRESCO survey were used to study the spatial distribution of star-forming regions and to revise slit-loss corrections for the Balmer fluxes. Our key findings are as follows:
\begin{itemize}
\item Paschen emission exhibits smaller half-light areas than rest-frame optical and near-IR continuum emission, and is distributed closer to the outskirts of galaxies. These results suggest that current star formation builds up stellar mass in more concentrated regions farther from the galaxy's center compared to the older stellar populations traced by the continuum, indicative of inside-out star formation. In galaxies where the nebular emission is significantly more concentrated than the continuum emission, slit-loss corrections will overestimate Balmer fluxes if they assume the nebular emission morphology follows the continuum.
\item The Balmer decrement and Paschen-based decrements (Pa$\alpha$/H$\alpha$ and Pa$\beta$/H$\alpha$) yield nebular reddening values that agree on average, but show substantial scatter among individual galaxies. Two galaxies exhibit significantly higher Paschen-derived reddening, which arises because non-unity dust covering fractions more significantly bias the Balmer decrement toward less reddened OB associations than the Paschen-based decrements, resulting in significantly higher Paschen reddening values when the bias is substantial enough.
\item Dust-corrected H$\alpha$ SFRs agree with dust-corrected Paschen-line SFRs on average, but four galaxies show significantly higher Paschen SFRs. As with nebular reddening, non-unity dust covering fractions bias H$\alpha$ SFRs toward lower values by weighting them toward less reddened OB associations, while Paschen SFRs trace star formation in the dustiest regions.
\item UV SFRs derived from best-fit SEDs assuming the SMC extinction curve to dust correct the stellar continuum align well with dust-corrected Paschen SFRs when averaged over all galaxies. Individually analyzed galaxies show significantly larger Paschen SFRs and lack increased scatter between Paschen and UV SFR at lower stellar masses, due to the $3\sigma$-detection requirement for Paschen lines, which biases the sample toward higher-SFR galaxies.
\item No significant correlations are found between O3N2 metallicity and $E(B-V)\textrm{-BD}$, or between $E(B-V)\textrm{-PB}$ and UV SFR. The lack of these expected correlations is consistent with non-unity dust covering fractions biasing the Balmer decrement and UV luminosities toward relatively unreddened OB associations, resulting in underestimated Balmer reddenings and UV SFRs.
    
\end{itemize}

Our study is limited by our reliance on a single Paschen line for the non-Balmer decrements and on slit-loss corrections for the Balmer fluxes. Future studies can address these limitations by employing more than one Paschen line to constrain nebular reddening and SFRs, using decrements that are independent of the Balmer lines and potentially more sensitive to star formation in the dustiest regions than those with only one Paschen line. Additionally, employing slitless spectroscopy for all lines or applying consistent slits for all measurements will mitigate systematic biases from flux corrections. Such studies can further evaluate the efficacy of Paschen lines relative to Balmer lines in tracing nebular reddening and SFRs in high-redshift galaxies, and can provide more robust corrections for inaccuracies in the measured Balmer reddening caused by non-unity dust covering fractions.

\section{Acknowledgments} \label{sec:Acknowledgments}

This work is based on observations made with the NASA/ESA/CSA James Webb Space Telescope. The data were obtained from the Mikulski Archive for Space Telescopes at the Space Telescope Science Institute, which is operated by the Association of Universities for Research in Astronomy, Inc., under NASA contract NAS5-03127 for \textit{JWST}. These observations are associated with Cycle 1 GO program \#1895. Support for program JWST-GO-1895 was provided by NASA through a grant from the Space Telescope Science Institute, under NASA contract NAS5-26555. We acknowledge support from NSF AAG grants AST1312780, 1312547, 1312764, and 1313171, grant AR13907 from the Space Telescope Science Institute, and grant NNX16AF54G from the NASA ADAP program. We thank the 3D-HST Collaboration, which provided the spectroscopic and photometric catalogs that were used to select the MOSDEF targets and derive stellar population parameters. This work has received funding from the Swiss State Secretariat for Education, Research and Innovation (SERI) under contract number MB22.00072, as well as from the Swiss National Science Foundation through project grant 200020\_207349. The Cosmic Dawn Center (DAWN) is funded by the Danish National Research Foundation under grant No.\ 140. IS acknowledges funding from the European Research Council (ERC) DistantDust (Grant No. 101117541) and the Atracc\'{i}on de Talento Grant No. 2022-T1/TIC-20472 of the Comunidad de Madrid, Spain. We wish to extend special thanks to those of Hawaiian ancestry on whose sacred mountain we are privileged to be guests. Without their generous hospitality, most of the observations presented herein would not have been possible.

\section{Data Availability} \label{sec:DataAvailability}

The MOSDEF data used here are publicly available and can be obtained at \hyperlink{https://mosdef.astro.berkeley.edu/for-scientists/data-releases/}{https://mosdef.astro.berkeley.edu/for-scientists/data-releases/}. The FRESCO observations used for the Paschen lines in this work are from public \textit{JWST} GO1 data (Programme \#1895), reduced with the publicly-available code \texttt{GRIZLI} (\url{grizli.readthedocs.io}). The reduced imaging data are publicly available at \url{https://s3.amazonaws.com/grizli-v2/JwstMosaics/v7/index.html} or through MAST: \url{https://archive.stsci.edu/hlsp/fresco} (DOI:10.17909/gdyc-7g80).

\newpage

\appendix
\section{Nebular Reddening Correlations} \label{sec:Appendix Correlations}

Table \ref{tab:appendix_table} shows the results of the statistical tests between $E(B-V) \textrm{-PB}$, $E(B-V) \textrm{-BD}$, and $\Delta E(B-V)$ and the galaxy properties analyzed in Section \ref{sec:Reddening Correlations}.

\begin{table}
\centering
\caption{}
\begin{tabular}{llcccc}
\toprule
Nebular Reddening & Galaxy Property & $\rm\rho_s$ \footnote{Spearman rank correlation coefficient} & $\rm P_S$ \footnote{Spearman p-value} & $\sigma$ \footnote{Confidence level that best-fit line has nonzero slope}\\
\midrule
\multirow{4}{*}{$E(B-V)\textrm{-PB}$} & $E(B-V)_\textrm{stars}$ & 0.365 & 0.013 & 7.004 \\
                         & $\log{(M_*)}$ & 0.374 & 0.010 & 5.587 \\
                         & $\rm 12+\log{(O/H)}$ & 0.518 & $8.72\times10^{-4}$ & 4.371 \\
                         & $\rm\log{(SFR\ [UV])}$ & 0.136 & 0.368 & 2.206 \\
                         
\midrule
\multirow{5}{*}{$E(B-V)\textrm{-BD}$} & $E(B-V)_\textrm{stars}$ & 0.352 & 0.016 & 2.673 \\
                         & $\log{(M_*)}$ & 0.479 & $7.68\times10^{-4}$ & 4.622 \\
                         & $\rm 12+\log{(O/H)}$ & 0.089 & 0.597 & 0.606 \\
                         & $\rm\log{(SFR\ [UV])}$ & 0.408 & $4.86\times10^{-3}$ & 4.035 \\
                         & $\rm\log{(SFR\ [Pa\alpha \ \& \ Pa\beta])}$ & 0.458 & $1.39\times10^{-3}$ & 4.142 \\
\midrule
\multirow{2}{*}{$\Delta E(B-V)$} & $\rm 12+\log{(O/H)}$ & 0.320 & 0.050 & 2.037 \\
                         & $\rm\log{(SFR\ [UV])}$ & -0.274 & 0.065 & 2.406 \\

\bottomrule
\end{tabular}
\label{tab:appendix_table}

\end{table}

\bibliography{bibliography}{}
\bibliographystyle{aasjournal}

\end{document}